\documentclass[a4paper,11pt]{article}
\pdfoutput=1
\usepackage{jheppub,booktabs}
\usepackage{amsmath,amssymb}
\usepackage{subfigure}
\usepackage{cancel}
\usepackage{mathrsfs}
\usepackage{graphicx}
\usepackage{hepunits}
\usepackage{wasysym}
\usepackage{feynmf}
\usepackage[utf8x]{inputenc}

\usepackage{indentfirst}
\usepackage{verbatim}
\usepackage{slashed}

\usepackage[percent]{overpic}

\usepackage{cancel}
\usepackage{mathrsfs}

\usepackage{multirow}

\newcommand{\Br}{\text{Br}}
\newcommand{\eq}[1]{(\ref{#1})}
\newcommand{\abs}[1]{\left|#1\right|}
\newcommand{\re}{\text{Re}}
\newcommand{\im}{\text{Im}}

\usepackage{tikz}
\usetikzlibrary{trees}
\usetikzlibrary{decorations.pathmorphing}
\usetikzlibrary{decorations.markings}

\graphicspath{{./Figs/}}

\title{Physical Constraints on a Class of Two-Higgs Doublet Models
with FCNC at tree level}

\author[a]{F. J. Botella,}
\author[b]{G. C. Branco,}
\author[c]{Adri\'an Carmona,}
\author[b]{M. Nebot,}
\author[b]{Leonardo Pedro,}
\author[b]{M. N. Rebelo}
\affiliation[a]{
Departament de F\'isica Te\`orica and IFIC,\\
Universitat de Val\`encia - CSIC, E-46100, Burjassot, Spain
}
\affiliation[b]{Centro de F\'isica Te\'orica de Part\'iculas, and Departamento de F\' \i sica\\
Instituto Superior T\'ecnico, Universidade de Lisboa, Av. Rovisco Pais, P-1049-001 Lisboa,
Portugal}
\affiliation[c]{
Institute for Theoretical Physics, \\
ETH Zurich, 8093 Zurich, Switzerland}

\emailAdd{fbotella@uv.es}
\emailAdd{gbranco@tecnico.ulisboa.pt}
\emailAdd{carmona@itp.phys.ethz.ch}
\emailAdd{nebot@cftp.ist.utl.pt}
\emailAdd{leonardo@cftp.ist.utl.pt}
\emailAdd{rebelo@tecnico.ulisboa.pt}
\preprint{IFIC/14-03,\ CFTP/14-004}
\abstract{We analyse the constraints and some of the phenomenological 
implications of a class of  two Higgs doublet models where
there are flavour-changing neutral currents (FCNC) at tree level but the potentially
dangerous FCNC couplings are suppressed by small entries of the CKM matrix $V$. This class
of models have the remarkable feature that, as a result of a discrete symmetry of the
Lagrangian, the FCNC couplings are entirely fixed in the quark sector by $V$
and the ratio $v_2 /v_1$  of the vevs of the neutral Higgs. 
The discrete symmetry is extended to the leptonic sector, so that there are FCNC
in the leptonic sector with their flavour structure fixed by the leptonic mixing matrix.
We analyse a large number
of processes, including decays mediated by charged Higgs at tree level, 
processes involving FCNC at tree level,  as well as loop induced processes.
We show that in this class of models one has new physical scalars beyond
the standard Higgs boson, with masses reachable at the next round of
experiments.}
\date{\today}
\begin{document}

\maketitle

\section{Introduction}
The recent discovery by both ATLAS \cite{ATLAS} and CMS \cite{CMS} of
a particle at about 125 GeV, which may be consistently interpreted as
a SM-like Higgs boson, has triggered an enormous interest in the
scalar sector of the SM and some of its extensions. A crucial
question to be probed experimentally is whether the scalar sector is
more complex than the one of the SM and in particular whether there
are more than one Higgs doublet. At least two Higgs doublets are
present in many extensions of the SM, in particular in some models
with spontaneous CP violation \cite{Lee:1973iz} and in supersymmetric extensions of
the SM. The general two Higgs doublet models (2HDM) \cite{Branco:2011iw,Djouadi:2005gj,Gunion:1989we} without extra
symmetries, have flavour changing neutral currents (FCNC) which have to be
suppressed in order to avoid conflict with experiment. The simplest way of
avoiding FCNC in the context of 2HDM is through the introduction of a
discrete symmetry leading to natural flavour conservation (NFC) \cite{Glashow:1976nt}. 
Another possibility of avoiding tree-level FCNC is through the hypothesis 
of aligned Yukawa couplings in flavour space \cite{Pich:2009sp}. 
Constraints arising from FCNC in the context of 2HDM have been the subject 
of many studies \cite{Hadeed:1985xn,Luke:1993cy,Cvetic:1998uw,Crivellin:2013wna,Mohapatra:2013cia}. A very interesting alternative to NFC is provided by the so-called BGL 
models \cite{Branco:1996bq,Botella:2009pq,Botella:2011ne}, where there are 
non-vanishing FCNC at tree level, but they are naturally suppressed as a
result of an exact symmetry of the Lagrangian, which is spontaneously
broken by the vevs of the neutral Higgs. The BGL models are highly constrained 
since, in the quark sector,
all couplings are fixed by $V$ and the ratio $v_2/v_1$ of the two vevs, 
with no other parameters. This is to be contrasted with the situation that one
encounters in the general 2HDM where there
is a large number of parameters which can be expressed in terms of
various unitary matrices arising from the misalignment in flavour
space between pairs of Hermitian flavour matrices \cite{Botella:2012ab}. The search for the allowed parameter space in two Higgs doublet models
has been done in the literature for a variety of scenarios
\cite{Basso:2012st,Cheon:2012rh,Altmannshofer:2012ar,Celis:2013rcs,Barroso:2013zxa,Grinstein:2013npa,Eberhardt:2013uba,Craig:2013hca,Ferreira:2013qua,Chang:2013ona,Celis:2013ixa,Harlander:2013qxa}. The extension
of BGL models to the leptonic sector is essential in order to allow
for the study of their phenomenological implications and, furthermore,
to allow for a consistent analysis of the renormalization group
evolution. The relationship between BGL-type models and the principle
of Minimal Flavour Violation (MFV) \cite{Buras:2000dm,D'Ambrosio:2002ex,Bobeth:2005ck,Dery:2013aba}
has been studied and a MFV
expansion was derived for the neutral Higgs couplings to fermions \cite{Botella:2009pq}.
In this paper, we analyse the constraints on BGL type models and discuss 
some of their phenomenological implications. 
This paper is organized as follows. In the next section, we briefly
review the BGL models and classify the various variants of these
models while at the same time settling the notation. In the third
section, we analyse the constraints on BGL models, derived from
experiment. In section \ref{SEC:Results} we present our results. The
explanation of the profile likelihood method used
in our analysis  and the input data appear in appendices. Finally, in section \ref{sec:concl}, 
we summarize our results and draw our conclusions.
   
\section{Theoretical framework}
We consider the extension of the SM consisting of the addition of two
Higgs doublets as well as three right-handed neutrinos. In this
work we only consider explicitly scenarios with Dirac type neutrinos,
where no Majorana mass terms are added to the Lagrangian. However, our 
analysis of the flavour-related experimental implications does not depend on the
nature of the neutrinos, i.e., Majorana or Dirac. Therefore, our
conclusions can be extended to the case of neutrinos being Majorana
fermions provided that deviations from unitarity of the $3 \times 3$
low energy leptonic mixing matrix are negligible, as it is the case in
most seesaw models. The extension of BGL models to the leptonic sector, both for Dirac and Majorana neutrinos, was addressed by some of the authors in \cite{Botella:2011ne}. 
In order to fix our notation, we explicitly write
the Yukawa interactions:
\begin{eqnarray}
{\mathcal{L}}_{Y} &=&-\overline{Q_{L}^{0}}\ \Gamma _{1}\Phi _{1}d_{R}^{0}-
\overline{Q_{L}^{0}}\ \Gamma _{2}\Phi _{2}d_{R}^{0}-\overline{Q_{L}^{0}}\
\Delta _{1}\tilde{\Phi }_{1}u_{R}^{0}-\overline{Q_{L}^{0}}\ \Delta _{2}
\tilde{\Phi }_{2}u_{R}^{0}  \nonumber \\
&&-\overline{L_{L}^{0}}\ \Pi _{1}\Phi _{1}\ell_{R}^{0}-\overline{L_{L}^{0}}\
\Pi _{2}\Phi _{2}\ell_{R}^{0}-\overline{L_{L}^{0}}\ \Sigma _{1}\tilde{\Phi} _{1}
\nu_{R}^{0}-\overline{L_{L}^{0}}\ \Sigma_{2}\tilde{\Phi}_{2}\nu_{R}^{0}+\mbox{h.c.}, \label{YukawaDirac1}
\end{eqnarray}
where $\Gamma_i$, $\Delta_i$  $ \Pi_{i}$ and $\Sigma_{i}$ are matrices in flavour space.

The quark mass matrices generated after spontaneous gauge symmetry
breaking are given by:
\begin{equation}
M_d = \frac{1}{\sqrt{2}} ( v_1  \Gamma_1 +
                           v_2 e^{i \theta} \Gamma_2 ), \quad 
M_u = \frac{1}{\sqrt{2}} ( v_1  \Delta_1 +
                           v_2 e^{-i \theta} \Delta_2 ),
\label{mmmm}
\end{equation}
where $v_i / \sqrt{2} \equiv |\langle 0|\phi^0_i|0\rangle|$ and $\theta$ denotes
the relative phase of the vacuum expectation values (vevs) of the
neutral components $\phi^0_i$ of $\Phi_i$. The matrices $M_d, M_u$ are
diagonalized by the usual bi-unitary transformations:
\begin{eqnarray}
U^\dagger_{dL} M_d U_{dR} = D_d \equiv \text{diag}\ (m_d, m_s, m_b) 
\label{umu}\,,\\
U^\dagger_{uL} M_u U_{uR} = D_u \equiv \text{diag}\ (m_u, m_c, m_t)\,.
\label{uct}
\end{eqnarray}
The neutral and the charged Higgs interactions obtained from the quark
sector of Eq.~(\ref{YukawaDirac1}) are of the form:
\begin{eqnarray}
{\mathcal L}_Y (\mbox{quark, Higgs})& = & - \overline{d_L^0} \frac{1}{v}\,
[M_d H^0 + N_d^0 R + i N_d^0 I]\, d_R^0   \nonumber \\
&&- \overline{{u}_{L}^{0}} \frac{1}{v}\, [M_u H^0 + N_u^0 R + i N_u^0 I] \,
u_R^{0}   \label{rep}\\
& & - \frac{\sqrt{2} H^+}{v} (\overline{{u}_{L}^{0}} N_d^0  \,  d_R^0 
- \overline{{u}_{R}^{0}} {N_u^0}^\dagger \,    d_L^0 ) + \text{h.c.} \nonumber 
\end{eqnarray}
where $v \equiv \sqrt{v_1^2 + v_2^2} $, and $H^0$, $R$ are orthogonal
combinations of the fields  $\rho_j$, arising when one expands
\cite{Lee:1973iz}  the neutral scalar fields around their vacuum
expectation values,
$\phi^0_j=\frac{e^{i\theta_j}}{\sqrt{2}}(v_j+\rho_j+i\eta_j)$,
choosing $H^0$ in such a way that it has couplings to the quarks which
are proportional to the mass matrices, as can be seen from
Eq.~(\ref{rep}). The required rotation is given by Eq.~(\ref{beta}).
Similarly, $I$ denotes the linear combination of
$\eta_{j}$ orthogonal to the neutral Goldstone boson. The matrices
$N_d^0$ and $N_u^0$ are given by:
\begin{equation}
N_d^0 = \frac{1}{\sqrt{2}} ( v_2  \Gamma_1 -
                           v_1 e^{i \theta} \Gamma_2 ), \quad 
N_u^0 = \frac{1}{\sqrt{2}} ( v_2  \Delta_1 -
                           v_1 e^{-i \theta} \Delta_2 ).
\end{equation}
In terms of the quark mass eigenstates $u, d$, the Yukawa couplings
are:
\begin{multline}
{\mathcal L}_Y (\mbox{quark, Higgs} ) =\\  - \frac{\sqrt{2} H^+}{v} \bar{u} \left(
V N_d \gamma_R - N^\dagger_u \ V \gamma_L \right) d +  \mbox{h.c.} - 
\frac{H^0}{v} \left(  \bar{u} D_u u + \bar{d} D_d \ d \right) -  \\
 -  \frac{R}{v} \left[\bar{u}(N_u \gamma_R + N^\dagger_u \gamma_L)u+
\bar{d}(N_d \gamma_R + N^\dagger_d \gamma_L)\ d \right] + \\
 +  i  \frac{I}{v}  \left[\bar{u}(N_u \gamma_R - N^\dagger_u \gamma_L)u-
\bar{d}(N_d \gamma_R - N^\dagger_d \gamma_L)\ d \right]
\end{multline}
where $\gamma_{L}$ and $\gamma_{R}$ are the left-handed and right-handed chirality projectors, respectively, and 	$N_d \equiv  U^\dagger_{dL} N_d^0 U_{dR}$, $N_u \equiv
U^\dagger_{uL} N_u^0 U_{uR}$, $V\equiv U^\dagger_{uL}U_{dL}$.

The flavour structure of the quark sector of two Higgs doublet models
is characterized by the four matrices $M_d$, $M_u$, $N_d^0$,
$N_u^0$. For the leptonic sector we have the corresponding matrices
which we denote by $M_\ell$, $M_\nu$, $N_\ell^0$, $N_\nu^0$.

In order to obtain a structure for $\Gamma _{i}$,  $\Delta _{i}$ such
that there are FCNC at tree level with strength completely controlled
by the Cabibbo -- Kobayashi -- Maskawa (CKM) mixing matrix $V$, Branco, Grimus and Lavoura (BGL) imposed the following
symmetry on the quark and scalar sector of the
Lagrangian \cite{Branco:1996bq}:
\begin{equation}
Q_{Lj}^{0}\rightarrow \exp {(i\tau )}\ Q_{Lj}^{0}\ ,\qquad
u_{Rj}^{0}\rightarrow \exp {(i2\tau )}u_{Rj}^{0}\ ,\qquad \Phi
_{2}\rightarrow \exp {(i\tau )}\Phi_{2}\ ,  \label{S symetry up quarks}
\end{equation}
where $\tau \neq 0, \pi$, with all other quark fields transforming 
trivially under the symmetry. The index $j$ can be fixed as either 1,
2 or 3. Alternatively the symmetry may be chosen as:
\begin{equation}
Q_{Lj}^{0}\rightarrow \exp {(i\tau )}\ Q_{Lj}^{0}\ ,\qquad
d_{Rj}^{0}\rightarrow \exp {(i2\tau )}d_{Rj}^{0}\ ,\quad \Phi
_{2}\rightarrow \exp {(- i \tau)}\Phi_{2}\ .  \label{S symetry down quarks}
\end{equation} 
The symmetry given by Eq.~(\ref{S symetry up quarks}) leads to Higgs
FCNC in the down sector, whereas the symmetry specified by 
Eq.~(\ref{S symetry down  quarks}) leads to Higgs FCNC in the up
sector. These two alternative choices of symmetry combined with the
three possible ways of fixing  the index $j$ give rise to six
different realizations of 2HDM with the flavour structure, in the
quark sector, controlled by the CKM matrix. 

In the leptonic sector, with Dirac type neutrinos, there is  perfect
analogy with the quark sector. The  requirement  that FCNC at tree
level  have  strength completely controlled by the Pontecorvo -- Maki
-- Nakagawa -- Sakata (PMNS) matrix, $U$ is enforced by one of the
following symmetries. Either 
\begin{equation}
L_{Lk}^{0}\rightarrow \exp {(i\tau )}\ L_{Lk}^{0}\ ,\qquad \nu
_{Rk}^{0}\rightarrow \exp {(i2\tau )}\nu _{Rk}^{0}\ ,\qquad \Phi
_{2}\rightarrow \exp {(i\tau )}\Phi _{2} \ , \label{S symetry neutrinos}
\end{equation}
or
\begin{equation}
L_{Lk}^{0}\rightarrow \exp {(i\tau )}\ L_{Lk}^{0}\ ,\qquad
\ell_{Rk}^{0}\rightarrow \exp {(i2\tau )}\ell_{Rk}^{0}\ ,\qquad \Phi
_{2}\rightarrow \exp {(-i \tau )}\Phi _{2} \ , \label{S symetry charged leptons}
\end{equation}
where, once again,  $\tau \neq 0, \pi$, with all other leptonic fields
transforming trivially under the symmetry. The index $k$ can be fixed
as either 1, 2 or 3. 

These are the so-called BGL type models that we analyse in this
paper. There are thirty six different  models corresponding to the
combinations of the six possible different implementations in each
sector\footnote{For Majorana neutrinos, implementing the symmetry reduces the number of models to eighteen possibilities, since in this case models with FCNC in the neutrino sector are not allowed.}. It is clear that in order to combine  the symmetry given by
Eq.~(\ref{S symetry up quarks}) with the one given by 
Eq.~(\ref{S symetry charged leptons}) an overall change of sign is
required, in one set of transformations. 

The symmetry given by Eq.~(\ref{S symetry up quarks})  with the choice
$j=3$ leads to the following pattern of zero textures for the Yukawa
couplings:
\begin{eqnarray}
 \Gamma_1 & = & 
\left[\begin{array}{ccc}  
\times  & \times & \times \\
\times & \times &  \times \\
0 & 0 & 0 
\end{array}\right], \qquad
 \Gamma_2   =  
\left[\begin{array}{ccc}  
0 & 0 & 0  \\
0 & 0 & 0 \\
\times & \times &  \times 
\end{array}\right], \label{gam}\\
 \Delta_1  & = & 
\left[\begin{array}{ccc}  
\times  & \times & 0 \\
\times & \times &  0 \\
0 & 0 & 0 
\end{array}\right], \qquad 
 \Delta_2   =  \left[\begin{array}{ccc}  
0  & 0 & 0 \\
0 & 0 &  0 \\
0 & 0 & \times
\end{array}\right], \label{del}
\end{eqnarray}
where $\times$ denotes an arbitrary entry. As a result of this
symmetry the matrices  $N_d$, $N_u$ are of the
form \cite{Branco:1996bq}:
\begin{equation}
(N_d)_{ij} = \frac{v_2}{v_1} (D_d)_{ij} - 
\left( \frac{v_2}{v_1} +  \frac{v_1}{v_2}\right) 
(V^\dagger)_{i3} (V)_{3j} (D_d)_{jj}\,, \label{24}
\end{equation}
whereas
\begin{equation}
N_u = - \frac{v_1}{v_2} \mbox{diag} \ (0, 0, m_t) +  \frac{v_2}{v_1}
\mbox{diag} \ (m_u, m_c, 0)\,. \label{25}
\end{equation}
In these equations only one new parameter not present in the SM
appears, to wit, the ratio $v_2/ v_1$. It is the presence of the above
symmetry, which prevents the appearance of additional free parameters.
As a result, BGL models are very constrained but these  constraints
crucially depend on the variant of the BGL model considered. For
example with the choice $j=3$ leading to Eqs.~(\ref{gam}),
(\ref{del}), (\ref{24}), Higgs mediated FCNC are controlled by the
elements of the third row of $V$. This leads, in a natural way,
to a very strong suppression in the FCNC entering in the ``dangerous"
$\Delta S = 2$ processes contributing to $K^0 - \bar K^0$
transitions. Indeed, in this variant of BGL models, the couplings entering in
the tree level $\Delta S = 2$ transition are proportional to
$|V_{td}V^\ast_{ts}|$ leading to a $\lambda ^{10}$ suppression in the
Higgs mediated $\Delta S = 2$ transition, where $\lambda\approx 0.2$
denotes the Cabibbo parameter. With this strong suppression even 
light neutral Higgs, with masses of the order $10^2$ GeV are
allowed. This strong natural suppression makes this variant of BGL
models specially attractive. The neutral mass eigenstates are linear
combinations of the fields $H^0$, $R$ and $I$ with the mixing
parameters determined by the Higgs potential.

Equations (\ref{gam}) and (\ref{del}) are written in the weak basis (WB) where 
the symmetry is imposed. The six different BGL models can be fully
defined in a covariant way under WB transformations
\cite{Botella:2009pq} by
 \begin{eqnarray}
N^0_d = \frac{v_2}{v_1} M_d - \left( \frac{v_2}{v_1} +  
\frac{v_1}{v_2}\right)    \mathcal{P}_{j}^{\gamma }\ M_d \,,
\label{bgl1} \\
N^0_u = \frac{v_2}{v_1} M_u - 
\left( \frac{v_2}{v_1} +  \frac{v_1}{v_2}\right)  
 \mathcal{P}_{j}^{\gamma } \ M_u\,, \label{bgl2}
\end{eqnarray}
together with
\begin{eqnarray}
\mathcal{P}_{j}^{\gamma }\Gamma _{2} &=&\Gamma _{2}\ ,\qquad \mathcal{P}
_{j}^{\gamma }\Gamma _{1}=0\ ,  \label{g1g2ge} \\
\mathcal{P}_{j}^{\gamma }\Delta _{2} &=&\Delta _{2}\ ,\qquad \mathcal{P}
_{j}^{\gamma }\Delta _{1}=0\ ,  \label{d1d2ge}
\end{eqnarray}
where $\gamma $ stands for $u$ (up) or $d$ (down) quarks, and 
$\mathcal{P}_{j}^{\gamma }$ are the projection operators
defined \cite{Botella:2004ks} by
\begin{equation}
\mathcal{P}_{j}^{u}=U_{uL}P_{j}U_{uL}^{\dagger }\ ,\qquad \mathcal{P}
_{j}^{d}=U_{dL}P_{j}U_{dL}^{\dagger }\ ,  \label{projectors1}
\end{equation}
and $\left( P_{j}\right) _{lk}=\delta _{jl}\delta _{jk}$. Obviously,
the zero textures written in the example given above only appear in
the special WB chosen by the symmetry. A change of WB will alter these
matrices without changing the physics. This fact leads to the
consideration of WB invariant conditions as a powerful tool to analyse
the physical implications of the flavour structure of models with two
Higgs doublets~\cite{Botella:2012ab}. The BGL example given explicitly
above corresponds to 
$\mathcal{P}_{j}^{\gamma } = \mathcal{P}_{3}^u\equiv U_{uL}P_3
U^\dagger_{uL} $.

With this notation the index $\gamma$ refers to the sector that has no
FCNC and $j$ refers to the row/column  of $V$ that parametrizes
the FCNC. Notice that for $\gamma$ denoting  ``up" the index $j$
singles a row of $V$, while for $\gamma$ denoting  ``down" the
index $j$ singles a column of $V$. A characteristic feature of
BGL models is the fact that in the WB covariant definition given by
Eqs.~(\ref{bgl1}) and (\ref{bgl2})  both matrices $N^0_d$,
$N^0_u$ involve the same projection operator. Different models with MFV
were obtained through the generalization of BGL models \cite{Botella:2009pq}.
Relaxing the above condition
allows, for instance, to build models with Higgs mediated FCNC in both
up and down sectors.  It has been argued that out of the models
verifying Eqs.~(\ref{g1g2ge}) and (\ref{d1d2ge}) and their
generalization to the leptonic sector, only BGL type models can be
enforced by some symmetry \cite{Botella:2011ne}. Furthermore, in
Ref.~\cite{Ferreira:2010ir} it was shown that BGL models are the only
models of this type that can be  enforced by abelian symmetries.
 
Similarly, for the leptonic sector, the symmetries of 
Eqs.~(\ref{S symetry neutrinos}) or, in alternative 
(\ref{S symetry charged leptons}), imply
\begin{eqnarray}
\mathcal{P}_{k}^{\beta }\Pi _{2} &=&\Pi _{2}\ ,\qquad \mathcal{P}_{k}^{\beta
}\Pi _{1}=0 \ , \label{P1,P2} \\
\mathcal{P}_{k}^{\beta }\Sigma _{2} &=&\Sigma _{2}\ ,\qquad \mathcal{P}
_{k}^{\beta }\Sigma _{1}=0 \ , \label{s1,s2}
\end{eqnarray}
where $\beta $ stands for neutrino ($\nu$) or for charged lepton ($\ell$)
respectively. In this case
\begin{equation}
\mathcal{P}_{k}^{\ell}=U_{\ell L}P_{k}U_{\ell L}^{\dagger }\ , \qquad 
\mathcal{P}_{k}^{\nu }=U_{\nu L}P_{k}U_{\nu L}^{\dagger }\ , \label{projectorsL}
\end{equation}
where $U_{\nu L}$ and $U_{\ell L}$ are the unitary matrices that
diagonalize the corresponding square mass matrices
\begin{eqnarray}
U_{\ell L}^{\dagger }M_{\ell}M_{\ell}^{\dagger }U_{\ell L} &=&\text{diag}\left( m_{e}^{2},m_{\mu
}^{2},m_{\tau }^{2}\right)\ ,  \nonumber \\
U_{\nu L}^{\dagger }M_{\nu }M_{\nu }^{\dagger }U_{\nu L} &=&\text{diag}\left(
m_{\nu _{1}}^{2},m_{\nu 2}^{2},m_{\nu 3}^{2}\right)\ ,
\label{massdiagonal lept}
\end{eqnarray}%
with $M_{\ell}$ and $M_{\nu}$ of the form
\begin{equation}
M_{\ell}=\frac{1}{\sqrt{2}}(v_{1}\Pi_{1}+v_{2}e^{i\theta}\Pi_{2})\ ,\quad 
M_{\nu }=\frac{1}{\sqrt{2}}(v_{1}\Sigma_{1}+v_{2}e^{-i\theta}\Sigma_{2})\ .  \label{massmatrixl1}
\end{equation}
In the leptonic sector, the PMNS mixing matrix
$U\equiv  U^\dagger_{\ell L}U_{\nu L}$, has large
mixings, unlike the CKM matrix $V$. Therefore, the Higgs mediated
FCNC are not strongly suppressed. However, models where the Higgs
mediated leptonic FCNC are present  only in the neutrino sector can be
easily accommodated experimentally due to the smallness of the neutrino masses. 

In the next sections we label each of the thirty six different models we
analyse by the pair ($\gamma_j$, $\beta_k$): the generation numbers $j,k$ refer to the projectors $P_{j,k}$
involved in each sector $\gamma,\beta$. For example, the model $(\text{up}_3, \ell_2)=(t,\mu)$  will have no tree level neutral flavour changing couplings in the up quark and the charged lepton sectors while the neutral flavour changing couplings in the down quark and neutrino sectors will be controlled, respectively, by $V_{td_i}^{\phantom{\ast}}V_{td_j}^\ast$ and $U_{\mu \nu_a}^{\phantom{\ast}}U_{\mu \nu_b}^\ast$.

In BGL models the Higgs potential is constrained by the imposed
symmetry to be of the form:
\begin{eqnarray}
V_\Phi&=&\mu_1 \Phi_1^{\dagger}\Phi_1+\mu_2\Phi_2^{\dagger}\Phi_2-\left(m_{12}\Phi_1^{\dagger}\Phi_2+\text{ h.c. }\right)+
2\lambda_3\left(\Phi^{\dagger}_1\Phi_1\right)\left(\Phi_2^{\dagger}\Phi_2\right)\nonumber\\
&+&2\lambda_4\left(\Phi_1^{\dagger}\Phi_2\right)\left(\Phi_2^{\dagger}\Phi_1\right)+
\lambda_1\left(\Phi_1^{\dagger}\Phi_1\right)^2+
\lambda_2\left(\Phi_2^{\dagger}\Phi_2\right)^2,
\end{eqnarray}
the term in $m_{12}$ is a soft symmetry breaking term. Its
introduction prevents the appearence of an would-be Goldstone boson
due to an accidental continuous global symmetry of the  potential,
which arises when the BGL symmetry is exact. Namely, in the limit
$m_{12} \rightarrow 0$  the pseudo scalar neutral field $I$ remains
massless. Hermiticity would allow the coefficient $m_{12}$ to be
complex, unlike the other coefficients of the scalar
potential. However,  freedom to rephase the scalar doublets allows to
choose without loss of generality all coefficients real. As a result,
$V_\Phi$ does not violate CP explicitly. It can also be easily shown that
it cannot violate CP spontaneously. In the absence of CP violation the
scalar field $I$ does not mix with the fields $R$ and $H^0$, therefore
$I$ is already a physical Higgs and the mixing of $R$ and $H^0$ is
parametrized by a single  angle.  There are  two important rotations 
that define the two parameters, $\tan \beta$ and $\alpha$, widely used in the literature:
\begin{eqnarray}
 \left( \begin{array}{c} H^0 \\ R  \end{array} \right) 
 =  \frac{1}{v}\left( \begin{array}{rr}  
v_1  & v_2 \\
- v_2 & v_1
\end{array} \right)
\left( \begin{array}{c} \rho_1\\ \rho_2  \end{array} \right)
 =  \left( \begin{array}{cc}  
\cos \beta  & \sin \beta \\
- \sin \beta & \cos \beta 
\end{array} \right)
\left( \begin{array}{c} \rho_1\\ \rho_2  \end{array} \right)
\label{beta}
\end{eqnarray}
This rotation ensures that the field $H^0$ has flavour conserving couplings to the quarks 
with strength equal to the standard model Higgs  couplings. The other rotation is:  
\begin{eqnarray}
 \left( \begin{array}{c} H \\ h  \end{array} \right) 
 =  \left( \begin{array}{cc}  
\cos \alpha  & \sin \alpha \\
- \sin \alpha & \cos \alpha
\end{array} \right)
\left( \begin{array}{c} \rho_1\\ \rho_2  \end{array} \right)
 \end{eqnarray}
relating  $\rho_1$ and $\rho_2$ to two of the neutral physical
Higgs  fields.
The seven  independent real parameters of the Higgs potential $V_\Phi$ will
fix the seven observable quantities, comprising the masses of the
three neutral Higgs, the mass of the charged Higgs,  the combination
$v \equiv \sqrt{v_1^2 + v_2^2} $, $\tan \beta \equiv v_2/v_1$, and $\alpha$.  In our
analysis we use the current limits on Higgs masses, identifying one of 
the Higgs with the one that was discovered by ATLAS and CMS. We make
the approximation of no mixing between $R$ and $H^0$ identifying $H^0$ with 
the recently discovered Higgs and $R$ and $I$ with the additional physical 
neutral Higgs fields. This limit corresponds to $\beta - \alpha = \pi/2 $ and
with this notation $H^0$ coincides with $h$, which is the usual choice in the
literature. This approximation is justified by the fact that
the observed Higgs boson seems to behave as a standard-like Higgs
particle. The quantity $v$ is of course already fixed by experiment.
Electroweak precision tests and, in particular the $T$ and $S$
parameters, lead to constraints relating  the masses of the new Higgs
fields among themselves.  Therefore the bounds on $T$ and $S$, together 
with  direct mass limits, significantly restrict the masses of the new Higgs particles, 
once  the mass of $H^\pm$ is fixed. In our analysis we study BGL type models by combining 
the six possible implementations of the quark sector with the six
implementations of the leptonic sector. It is illustrative to plot our
results in terms of $m_{H^{\pm}}$ versus $\tan \beta$, since, as explained above
in the context of our approximation of no mixing between $R$ and
$H^0$, there is not much freedom left. Therefore with these two
parameters we approximately scan the whole region of parameter
space. In our analysis, we impose present constraints from several
relevant flavour observables, as specified in the next section. 

\section{Confronting experimental results\label{SEC:ExpConst}}

\subsection{Generalities\label{SEC:ExpConst-sSEC:Gen}}
In the class of 2HDM considered in this paper, the Yukawa interactions of the new scalars may produce new contributions, at tree and at loop level, that modify the SM predictions for many processes for which experimental information is available. As is customary, this will allow us to study the viability and interest of the different cases within this class of models. In terms of the New Physics (NP) and the SM leading contributions, one can organize the processes to be considered as follows.
\begin{itemize}
\item Processes with tree level NP contributions mediated by $H^\pm$ and SM tree level contributions $W^\pm$-mediated, as, for example, universality in lepton decays, leptonic and semileptonic decays of mesons like $\pi\to e \nu$, $B\to\tau\nu$ and $B\to D\tau\nu$, or $\tau$ decays of type $\tau\to M\nu$.
\item Processes with tree level NP contributions mediated by the neutral scalars $R$, $I$, and
\begin{itemize}
\item loop level SM contributions as in, for example, $K_L\to\mu^+\mu^-$, $B_s\to\mu^+\mu^-$, and $B^0\rightleftarrows\bar B^0$ oscillations,
\item highly suppressed (because of the smallness of the neutrino masses) loop level SM contributions as in, for example, $\tau^-\to\mu^-\mu^-\mu^+$ or $\mu^-\to e^-e^-e^+$.
\end{itemize}
\item Processes with loop level NP contributions and
\begin{itemize}
\item loop level SM contributions as in, for example, $B\to X_s\gamma$,
\item highly suppressed (here too because of the smallness of the neutrino masses) loop level SM contributions as in, for example, $\tau\to\mu\gamma$ or $\mu\to e\gamma$.
\end{itemize}
\end{itemize}

Besides those observables, electroweak precision information -- $Z\to b\bar b$ and the oblique parameters $S$, $T$ -- are also relevant; they involve loop level contributions from the new scalars.

Table \ref{TAB:summary} summarizes this classification of the potentially relevant observables. Notice however that the table signals the possible new contributions but for each specific model type, some of them will be absent. More detailed descriptions of each type of constraint are addressed in the following subsections. Since we focus in the flavour sector, we exclude from the analysis of the experimental implications of the BGL models processes that probe additional couplings related to the scalar potential, such as $H^0\to\gamma\gamma$, central in the Higgs discovery at the LHC, and refer the interested reader to \cite{Bhattacharyya:2013rya}.

\newcommand{\nchck}{{\checkmark}}
\newcommand{\grchck}{{\footnotesize\color{gray}\nchck}}
\begin{table}[h] 
\begin{center}
\begin{tabular}{c|cc|cc|cc|}
\cline{2-7} & \multicolumn{4}{|c|}{BGL - 2HDM} & \multicolumn{2}{||c|}{SM}\\ 
\cline{2-7} & \multicolumn{2}{|c|}{Charged $H^\pm$} & \multicolumn{2}{|c|}{Neutral $R$, $I$} & \multicolumn{1}{||c|}{\multirow{2}{*}{Tree}} & \multicolumn{1}{|c|}{\multirow{2}{*}{Loop}}\\
\cline{2-5} & \multicolumn{1}{|c|}{Tree} & \multicolumn{1}{|c|}{Loop} & \multicolumn{1}{|c|}{Tree} & \multicolumn{1}{|c|}{Loop} & \multicolumn{1}{||c|}{} & \multicolumn{1}{|c|}{} \\
\hline\multicolumn{1}{|c|}{$M\to\ell\bar\nu,M^\prime\ell\bar\nu$} & \multicolumn{1}{|c|}{\nchck} & \grchck & \multicolumn{1}{|c|}{} & \multicolumn{1}{|c|}{\grchck} & \multicolumn{1}{||c|}{\nchck} & \multicolumn{1}{|c|}{\grchck}\\
\hline\multicolumn{1}{|c|}{Universality} & \multicolumn{1}{|c|}{\nchck} & \grchck & \multicolumn{1}{|c|}{} & \multicolumn{1}{|c|}{\grchck} & \multicolumn{1}{||c|}{\nchck} & \multicolumn{1}{|c|}{\grchck}\\
\hline\hline\multicolumn{1}{|c|}{$M^0\to\ell_1^+\ell_2^-$} & \multicolumn{1}{|c|}{} & \grchck & \multicolumn{1}{|c|}{\nchck} & \multicolumn{1}{|c|}{\grchck} & \multicolumn{1}{||c|}{} & \multicolumn{1}{|c|}{\nchck}\\
\hline\multicolumn{1}{|c|}{$M^0\rightleftarrows \bar M^0$} & \multicolumn{1}{|c|}{} & \grchck & \multicolumn{1}{|c|}{\nchck} & \multicolumn{1}{|c|}{\grchck} & \multicolumn{1}{||c|}{} & \multicolumn{1}{|c|}{\nchck}\\
\hline\multicolumn{1}{|c|}{$\ell_1^-\to\ell_2^-\ell_3^+\ell_4^-$} & \multicolumn{1}{|c|}{} & \grchck & \multicolumn{1}{|c|}{\nchck} & \multicolumn{1}{|c|}{\grchck} & \multicolumn{1}{||c|}{} & \multicolumn{1}{|c|}{\grchck}\\
\hline\hline\multicolumn{1}{|c|}{$B\to X_{s}\gamma$} & \multicolumn{1}{|c|}{} & \nchck & \multicolumn{1}{|c|}{} & \multicolumn{1}{|c|}{\nchck} & \multicolumn{1}{||c|}{} & \multicolumn{1}{|c|}{\nchck}\\
\hline\multicolumn{1}{|c|}{$\ell_j\to \ell_i\gamma$}  & \multicolumn{1}{|c|}{} & \nchck  & \multicolumn{1}{|c|}{} & \multicolumn{1}{|c|}{\nchck} & \multicolumn{1}{||c|}{} & \multicolumn{1}{|c|}{\grchck}\\
\hline\hline\multicolumn{1}{|c|}{EW Precision} & \multicolumn{1}{|c|}{} & \nchck & \multicolumn{1}{|c|}{} & \multicolumn{1}{|c|}{\nchck} & \multicolumn{1}{||c|}{} & \multicolumn{1}{|c|}{\nchck}\\
\hline
\end{tabular}
\end{center}
\caption{Summary table of the different types of relevant observables; leading contributions are tagged $\nchck$ while subleading or negligible ones are tagged $\grchck$.\label{TAB:summary}}
\end{table}

The set of observables that we consider is sufficient to obtain significant constraints for the masses of the new scalars and $\tan\beta$. Notice that, since the new contributions will be typically controlled by these masses, $\tan\beta$ and the mixing matrices, with no additional parameters, we need fewer observables than would be necessary in the analysis of a more general 2HDM such as the one presented in \cite{Crivellin:2013wna}. 

Apart from the previous flavour related  observables, direct searches at colliders may be relevant. For instance, a charged Higgs decaying to $\tau^+\nu$ or $c\bar{s}$ with a mass lighter than $80$ \GeV ~was excluded\footnote{For all BGL models, in the parameter space not excluded by the previous observables, the branching ratio for the decays $H^\pm \to \tau^+\nu$ and $H^\pm \to c\bar{s}$ is larger than $96\%$ and this bound applies.}, in the context of 2HDM, at LEP \cite{LEP}. 
However, we do not include recent results from searches at the LHC like \cite{Aad:2012tj} and \cite{Chatrchyan:2012vca} since: (a) a type II 2HDM is typically assumed, and thus such bounds are not directly valid for most BGL models (and the appropriate model specific analysis goes beyond the scope of this work), and (b) furthermore this allows us to show that there are BGL models where the flavour observables we are taking into consideration, by themselves, do not impose such stringent bounds and allow light charged Higgs masses which may be probed at colliders, in particular at the LHC.


In the next subsections we describe in detail the different types of observables introduced above.

\subsection{Processes mediated by charged scalars at tree level\label{SEC:ExpConst-sSEC:TreeCharged}}
Since transitions mediated within the SM by a $W$ boson may receive new $H^\pm$ mediated contributions, one has to pay attention to:
\begin{itemize}
\item universality tests in pure leptonic decays $\ell_1\to\ell_2\nu\bar\nu$,
\item leptonic decays of pseudoscalar mesons $M\to \ell\nu$,
\item semileptonic decays of pseudoscalar mesons $M\to M^\prime\ell\nu$,
\item $\tau$ decays of the form $\tau\to M\nu$.
\end{itemize}

\subsubsection{Universality}
 Pure leptonic decays $\ell_1\to\ell_2\nu\bar\nu$ are described by the following effective Lagrangian
\begin{multline}
{\mathcal L}_{\rm eff}=-\frac{4 G_F}{\sqrt{2}}\times \\ \sum_{\ell_\alpha,\ell_\beta=e,\mu,\tau}\sum_{i,j=1}^3 U^\ast_{\ell_\alpha\nu_i}U_{\ell_\beta \nu_j}
\left\{\left[\bar \nu_i\gamma^\mu \gamma_L \ell_\alpha\right] 
\left[\bar \ell_\beta\gamma_\mu\gamma_L\nu_j\right]+  g^{\nu_i\ell_\alpha \nu_j\ell_\beta}\left[\bar \nu_i\gamma_R \ell_\alpha\right] \left[\bar \ell_\beta\gamma_L\nu_j\right]\right\}.\label{eq:L:univ:lept}
\end{multline}
The second operator in \eq{eq:L:univ:lept} is the new contribution mediated by $H^\pm$. The coefficient $g^{\nu_i\ell_\alpha \nu_j\ell_\beta}$ depends on the specific BGL model:
\begin{equation}
g^{\nu_i\ell_\alpha \nu_j\ell_\beta}=-\frac{m_{\ell_\alpha} m_{\ell_\beta}}{m_{H^+}^2}C^{\nu_i\ell_\alpha}C^{ \nu_j\ell_\beta}\,,
\end{equation}
where, $C^{\nu_i\ell_\alpha}=-1/\tan\beta$ for models of types $\nu_i$ and $\ell_\alpha$ and $C^{\nu_i\ell_\alpha}=\tan\beta$ otherwise -- this concerns the lepton label of the model, the quark one is irrelevant here.
Following the notation in \cite{Pich:1995,Pich:2010}, we then have
\begin{align}
&\abs{g_{RR,\ell_\alpha \ell_\beta}^{S}}^2\equiv\sum_{i,j=1}^3|U_{\ell_\alpha \nu_i}|^2|U_{\ell_\beta \nu_j}|^2(g^{\nu_i\ell_\alpha \nu_j\ell_\beta})^2\,,\\
&\abs{g_{LL,\ell_\alpha \ell_\beta}^{V}}^2\equiv 1\,,\\
&\left(g_{RR,\ell_\alpha \ell_\beta}^{S}\right) \left(g_{LL,\ell_\alpha \ell_\beta}^{V}\right)^*\equiv\sum_{i,j=1}^3|U_{\ell_\alpha \nu_i}|^2|U_{\ell_\beta \nu_j}|^2 g^{\nu_i\ell_\alpha \nu_j\ell_\beta}\,.
\end{align}
We consider for example universality in $\tau$ decays,
\begin{equation}
\abs{\frac{g_\mu}{g_e}}^2\equiv \frac{\text{Br}\left(\tau\to \mu\nu\bar\nu\right)}{\text{Br}\left(\tau \to e\nu\bar\nu\right)}
\frac{f\big(\frac{m^2_e}{m^2_\tau}\big)}{f\big(\frac{m^2_\mu}{m^2_\tau}\big)}\,,
\end{equation}
where
\begin{equation}
\frac{\text{Br}(\tau\to \mu\nu\bar\nu)}{\text{Br}(\tau \to e\nu\bar\nu)}=
\frac{\left(\big|{g_{LL,\tau \mu}^{V}}\big|^2+\frac{1}{4}\big|{g_{RR,\tau \mu}^{S}}\big|^2\right)
f\big(\frac{m^2_\mu}{m^2_\tau}\big)+
2{\rm Re}\left(g_{RR,\tau \mu}^{S} \left(g_{LL,\tau \mu}^{V}\right)^\ast\right)
\frac{m^2_\mu}{m^2_\tau} g\big(\frac{m^2_\mu}{m^2_\tau}\big)}{
\left(\big|{g_{LL,\tau e}^{V}}\big|^2+\frac{1}{4}\big|{g_{RR,\tau e }^{S}}\big|^2\right)
f\big(\frac{m^2_e}{m^2_\tau}\big)+2{\rm Re}\left(g_{RR,\tau e }^{S}
\left(g_{LL,\tau e }^{V}\right)^\ast\right)\frac{m^2_e}{m^2_\tau} g\big(\frac{m^2_e}{m^2_\tau}\big)}\,,\label{eq:univ:BRs}
\end{equation}
with $f(x)$ and $g(x)$ phase space functions\footnote{$f(x)=1-8x+8x^3-x^4-12x^2 \log(x)$ and $g(x)=1+9x-9x^2-x^3+6x(1+x)\log(x)$.}. One loop radiative corrections for the individual branching ratios cancel out in the ratio \eq{eq:univ:BRs}. The experimental limits on $\abs{g_{RR,\ell_\alpha\ell_\beta}^{S}}$ are collected in appendix \ref{AP:Input}.

\subsubsection{Semileptonic processes}
 Semileptonic processes may also receive tree level contributions from virtual $H^\pm$; the relevant effective Lagrangian for these processes is:
\begin{multline}
{\mathcal L}_{\rm eff} = -\frac{4 G_F}{\sqrt{2}}\ \sum_{u_i=u,c,t}\ \sum_{d_j=d,s,b}\ \sum_{\ell_a=e,\mu,\tau}\,\sum_{\nu_b=\nu_1,\nu_2,\nu_3}\ V_{u_id_j}\ U_{\ell_a \nu_b}\\
\left\{\left[\bar u_i\gamma^\mu \gamma_L d_j\right]\left[\bar \ell_a\gamma_\mu\gamma_L\nu_{b}\right] + \left[\bar u_i \left( g_L^{u_id_j \nu_b\ell_a}\,\gamma_L + g_R^{u_id_j \nu_b\ell_a}\, 
\gamma_R\right) d_j\right] \left[\bar \ell_a\gamma_L\nu_b\right]\right\}+ \mbox{h.c.}\,,\label{eq:L:semilept}
\end{multline}
where
\begin{equation}
g_L^{u_id_j \nu_b\ell_a}=\frac{ m_{u_i} m_{\ell_a}}{m_{H^+}^2}C^{u_id_j}C^{\ell_a \nu_b}\,,\qquad g_R^{u_id_j \nu_b\ell_a}=-\frac{ m_{d_j} m_{\ell_a}}{m_{H^+}^2}C^{u_id_j}C^{\nu_b\ell_a},
\end{equation}
and, $C^{u_id_j}=-1/\tan\beta$ for models of types $u_i$ and $d_j$, $C^{u_id_j}=\tan\beta$ otherwise, while $C^{ \nu_b\ell_a}=-1/\tan\beta$ for models of types $\ell_a$ and $\nu_b$, $C^{\nu_b\ell_a }=\tan\beta$ otherwise.

\begin{figure}[h]
\begin{center}
\subfigure[$M\to\ell\nu$]{\raisebox{40pt}{\includegraphics[width=0.32\textwidth]{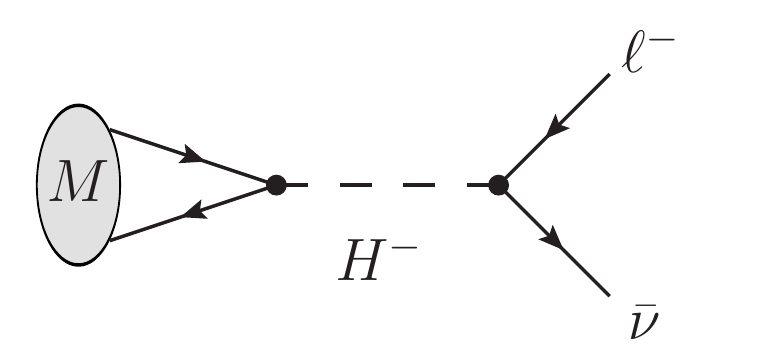}}}\quad
\subfigure[$M\to M^\prime\ell\nu$]{\includegraphics[width=0.3\textwidth]{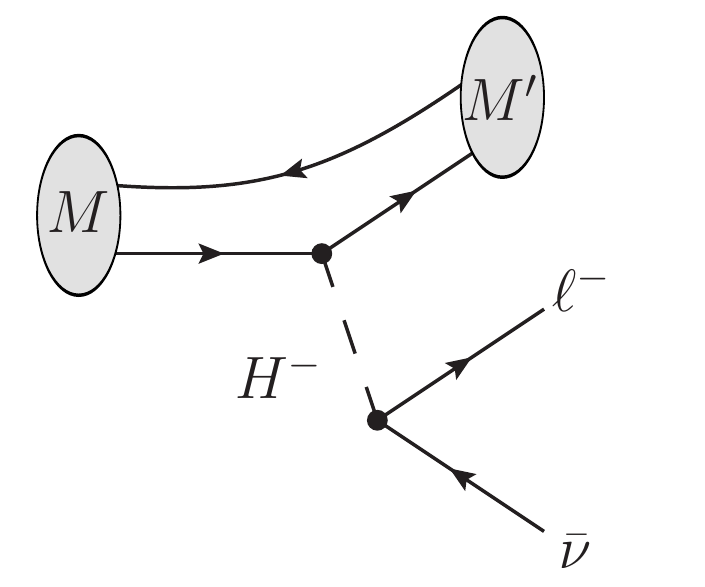}}\quad
\subfigure[$\tau\to M\nu$]{\raisebox{20pt}{\includegraphics[width=0.32\textwidth]{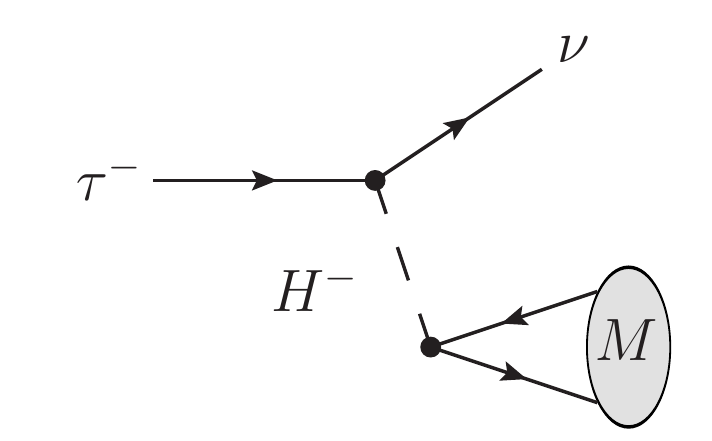}}}
\caption{Tree level $H^\pm$ mediated NP contributions to semileptonic process.\label{fig:Semileptonic}}
\end{center}
\end{figure}
 The rate of the leptonic decay $M\to\ell\bar\nu$ of a pseudoscalar meson $M$, with quark content $\bar u_i d_j$, obtained from the effective Lagrangian in Eq.~\eq{eq:L:semilept}, is given by\footnote{Including electromagnetic radiative corrections 
\cite{kaons}, $\Gamma(M\to \ell\bar\nu)=(1+\delta_{\rm em})\,\Gamma_0(M\to \ell\bar\nu)$.}
\begin{equation}
\Gamma_0(M\to \ell\bar\nu)\, =\, G_F^2m_\ell^2 f_M^2 |V_{u_id_j}|^2 \,\frac{m_{M}}{8\pi} \left( 1- \frac{m_\ell^2}{m_{M}^2}\right)^2\; 
\sum_{n=1,2,3}|U_{\ell \nu_n}|^2|1-\Delta^{ \nu_n\ell}_{u_id_j}|^2\,. \label{eq:semileptonic:rate0:Mln}
\end{equation}
The scalar mediated new contribution is given by,
\begin{equation}
\Delta^{ \nu_n\ell}_{u_id_j}=C^{u_id_j}C^{\nu_n\ell}\frac{m_M^2}{m_{H^\pm}^2}\,.
\end{equation}
Since the process is helicity suppressed and receives NP contributions proportional to $m_M^2/m_{H^{\pm}}^2$, interesting channels are expected to involve heavy mesons and the $\tau$ lepton, as for example in $B^+\to\tau^+\nu$, $D_s^+\to\tau^+\nu$.
Taking into account the different possible values of $C^{u_id_j}$ and $C^{\nu_n\ell}$, we must have
\[
C^{u_id_j}C^{\nu_n\ell}\in\left\{-1,\ \tan^2\beta,\ \frac{1}{\tan^2\beta}\right\}\,.
\]
Therefore, for $m_{H^+}^2\gg m^2_M$, if $\Delta_{u_id_j}^{\nu_n\ell}$ is negative, then the NP contribution is negligible; otherwise, if the NP contribution is enhanced by $(\tan\beta)^{\pm 2}$, it will typically interfere destructively with the SM contribution. An increase with respect to SM predictions, which would be interesting for example to account for some $B^+\to\tau^+\nu$ measurements, would require a NP contribution more than twice larger than the SM one, leading to tensions in other observables. The different channels considered in the analysis are collected in appendix \ref{AP:Input} and radiative corrections are included according to \cite{kaons}.

 In the case of $\tau$ decays of type $\tau\to M\nu$, the analogue of Eq.~(\ref{eq:semileptonic:rate0:Mln}) is\footnote{Radiative corrections to $\Gamma_0(\tau\to M\nu)$ are included in the analysis \cite{taucorrections}.} 
\begin{equation}
\Gamma_0(\tau\to M\nu)\, =\, G_F^2m_\tau^3 f_M^2 |V_{u_id_j}|^2 \,\frac{3}{16\pi} \left( 1- \frac{m_{M}^2}{m_\tau^2}\right)^2\; 
\sum_{n=1,2,3}|U_{\tau \nu_n}|^2|1-\Delta^{\nu_n\tau}_{u_id_j}|^2\,. \label{eq:semileptonic:rate0:tMn}
\end{equation}
The analysis uses experimental $\tau\to\pi\nu$ and $\tau\to K\nu$ results -- see table \ref{TAB:AP:TreeCharged}.

 While $M\to\ell\bar\nu$ transitions are helicity suppressed two body decays, this is not the case anymore for $M\to M^\prime \ell\bar\nu$ decays. The corresponding decay amplitude is described by two form factors, $F_+(q^2)$ and $F_0(q^2)$ -- with $q$ the momentum transfer to the $\ell\bar\nu$ pair --, associated to the P wave and the S wave components of the amplitude $\langle 0 | \bar u_i\gamma^\mu d_j| M \bar M^\prime \rangle$. The $H^\pm$ mediated amplitude can only contribute to the S wave component. Considering for example a specific case like $B\to D\tau\nu$, where the quark level weak transition is $b\to c\tau\nu$, we have
\begin{equation}
	\frac{F_0^{(\mathrm{BGL})}(q^2,n)}{F_0^{(\mathrm{SM})}(q^2)}= 1-C^{cb}C^{\nu_n\tau}\frac{q^2}{m_{H^+}^2}\,,
\end{equation}
giving then
\begin{multline}
	\frac{\Gamma_{(\mathrm{BGL})}(B\to D \tau\nu)}{\Gamma_{(\mathrm{SM})}(B\to D \tau\nu)}=1+\\
\sum_{n=1}^3|U_{\tau\nu_n}|^2 \left(-{C_1} C^{cb}C^{\nu_n\tau}\frac{m_\tau(m_b-m_c)}{m_{H^+}^{2}}
+C_2(C^{cb}C^{\nu_n\tau})^2\frac{ m_\tau^{2}(m_b-m_c)^{2}}{ m_{H^+}^{4}}\right)\,,
\end{multline}
with coefficients $C_1\sim 1.5$ and $C_2\sim 1.0$. For $B\to D^\ast \tau\nu$, we have instead 
\begin{multline}
	\frac{\Gamma_{(\mathrm{BGL})}(B\to D^\ast \tau\nu)}{\Gamma_{(\mathrm{SM})}(B\to D^\ast \tau\nu)}=1+\\
\sum_{n=1}^3|U_{\tau\nu_n}|^2 \left(-{C_1} C^{cb}C^{\nu_n\tau}\frac{m_\tau(m_b+m_c)}{M_{H^+}^{2}}
+C_2(C^{cb}C^{\nu_n\tau})^2\frac{ m_\tau^{2}(m_b+m_c)^{2}}{ M_{H^+}^{4}}\right)\,,
\end{multline}
and $C_1\sim 0.12$ and $C_2\sim 0.05$. 
Notice that, even though BGL models still remain compatible with the present data
for the decays $B \to \tau \nu$, $B\to D \tau \nu$ and $B\to D^\ast \tau\nu$,
if the experimental anomalies observed in these processes, pointing towards
physics beyond the SM, are confirmed no two such anomalies could be simultaneously
accommodated in the BGL framework.

For $K\to\pi\ell\nu$ decays, rather than resorting to the rate or the branching fraction to constrain the NP contributions, the Callan-Treiman relation is used to relate the scalar form factor at the kinematic point $q^2_{\rm CT}=m_K^2-m_\pi^2$ to the decay constants of $K$ and $\pi$:
\begin{equation}
\frac{F_0^{(BGL)}(q^2_{\rm CT})}{F_+(0)}=\frac{f_K}{f_\pi}\frac{1}{F_+(0)}+\Delta_{\chi {\rm PT}}\equiv C\,.\label{eq:CT:KPln}
\end{equation}
$\Delta_{\chi {\rm PT}}$ is a Chiral Perturbation Theory correction. The right-hand side of Eq.~(\ref{eq:CT:KPln}), $C$, is extracted from experiment,  thus leading
to a constraint on $F_0^{(\mathrm{BGL})}(q^2_{\rm CT})$.

\subsection{Processes mediated by neutral scalars at tree level\label{SEC:ExpConst-sSEC:TreeNeutral}}
While the $H^\pm$ mediated NP contributions of the previous section compete with tree level SM amplitudes -- including suppressed ones, as in $M\to\ell\nu$ decays --, the neutral scalars $R$ and $I$ produce tree level contributions that compete with loop level SM contributions. We consider three different types of processes.
\begin{itemize}
\item Lepton flavour violating decays $\ell_1^-\to\ell_2^-\ell_3^+\ell_4^-$: in this case the SM loop contribution, proportional to neutrino masses is completely negligible and thus NP provides the only relevant one.
\item Mixings of neutral mesons, $M^0\rightleftarrows \bar M^0$, where $M^0$ could be a down-type meson $K^0$, $B^0_d$ or $B^0_s$ or the up-type meson $D^0$. The distinction among down and up-type mesons is relevant since depending on the BGL model the tree level NP contributions will appear in one or the other sector.
\item Rare decays $M^0\to\ell_1^+\ell_2^-$ (including lepton flavour violating modes $\ell_1\neq\ell_2$): again depending on the BGL model and $M^0$ being one of the previous down or up-type pseudoscalar mesons, the tree level NP contributions will be present or not. 
\end{itemize}
 
\subsubsection{Lepton flavour violating decays }
 
\begin{figure}[h]
\begin{center}
\includegraphics[width=0.5\textwidth]{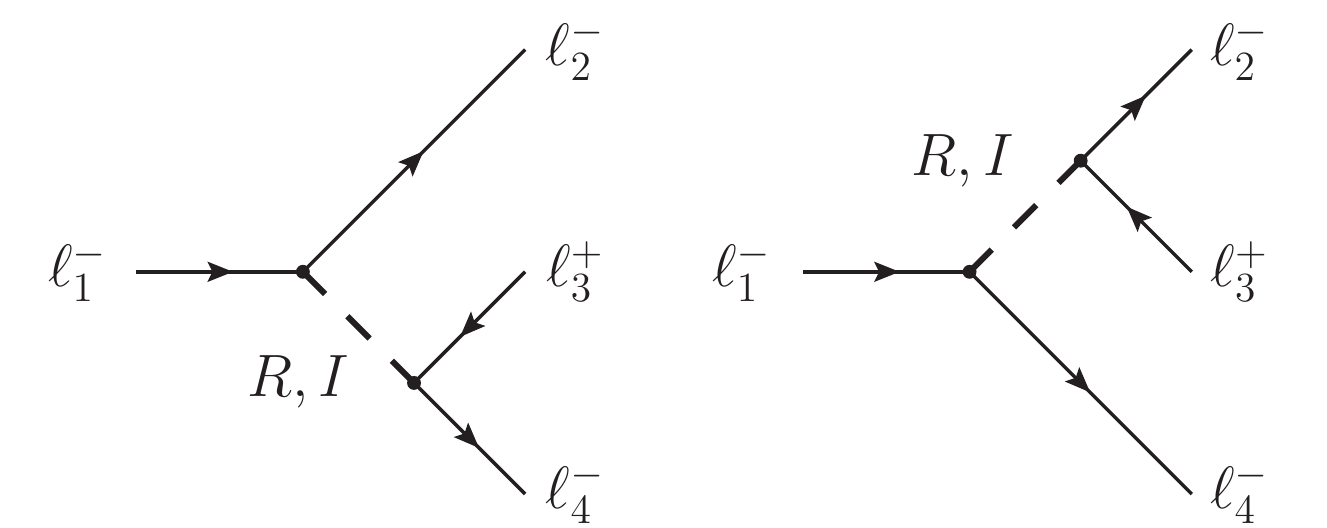}
\caption{Tree level $R,I$ mediated NP contributions to $\ell_1^-\to \ell_2^-\ell_3^+\ell_4^-$.\label{fig:Mtoll}}
\end{center}
\end{figure}

Lepton flavour violating decays of the form $\ell_1^-\to\ell_2^-\ell_3^+\ell_4^-$,
such as $\mu^-\to e^-e^+e^-$, $\tau^-\to e^-\mu^+\mu^-$ or $\tau^-\to \mu^- e^+ \mu^-$
are completely negligible in the SM, since the corresponding penguin and/or box amplitudes
are proportional to neutrino masses. In BGL models of type $(X,\nu_j)$, tree level
NP contributions mediate these decays. For muons, there is only one possible decay
of this type, while for taus there are two interesting cases: either $\ell_3^+$
belongs to the same family as one of the negatively charged leptons or not.
In the latter case the two vertices in the diagrams of figure \ref{fig:Mtoll} are flavour changing
and the SM contributes dominantly via a box diagram. Otherwise, the dominant BGL
contribution only requires one flavour changing vertex and SM penguin diagrams are
possible. In this case a connection can be established with the lepton flavour
violating processes of the type $\ell_j \to \ell_i \gamma$ considered in section \ref{SEC:ExpConst-sSEC:Loop}.

The corresponding effective Lagrangian is
\begin{multline}
\mathcal L_{\rm eff} = 
-\frac{2 G_F}{\sqrt{2}}\sum_{\chi_1,\chi_2=L,R}
\Big\{
g_{\chi_1\chi_2}^{12,34} \left[\bar\ell_2\gamma_{\chi_1} \ell_1\right]\left[\bar\ell_4\gamma_{\chi_2} \ell_3\right]+
g_{\chi_1\chi_2}^{14,32} \left[\bar\ell_4\gamma_{\chi_1} \ell_1\right]\left[\bar\ell_2\gamma_{\chi_2} \ell_3\right]
\Big\}\,,
\end{multline}
with
\begin{align*}
&g_{LL}^{ij,kl}=\frac{(N_\ell^\dagger)_{\ell_j \ell_i}(N_\ell^\dagger)_{\ell_l \ell_k}}{m_R^2}-\frac{(N_\ell^\dagger)_{\ell_j \ell_i}(N_\ell^\dagger)_{\ell_l \ell_k}}{m_I^2}\,,\ &
g_{RL}^{ij,kl}=\frac{(N_\ell)_{\ell_j \ell_i}(N_\ell^\dagger)_{\ell_l \ell_k}}{m_R^2}+\frac{(N_\ell)_{\ell_j \ell_i}(N_\ell^\dagger)_{\ell_l \ell_k}}{m_I^2}\,,\\
&g_{LR}^{ij,kl}=\frac{(N_\ell^\dagger)_{\ell_j \ell_i}(N_\ell)_{\ell_l \ell_k}}{m_R^2}+\frac{(N_\ell^\dagger)_{\ell_j \ell_i}(N_\ell)_{\ell_l \ell_k}}{m_I^2}\,,\ &
g_{RR}^{ij,kl}=\frac{(N_\ell)_{\ell_j \ell_i}(N_\ell)_{\ell_l \ell_k}}{m_R^2}-\frac{(N_\ell)_{\ell_j \ell_i}(N_\ell)_{\ell_l \ell_k}}{m_I^2}\,,
\end{align*}
and $N_\ell$ is the analogue, in the lepton sector, of $N_d$, i.e. the analogue of Eq.~(\ref{bgl1}) in the basis where $M_\ell$ is diagonal. Neglecting all masses except $m_{\ell_1}$, the width of the process is\footnote{The factor $(1+\delta_{\ell_2\ell_4})^{-1}$ takes into account the case of two identical particles in the final state.}
\begin{multline}
\Gamma(\ell_1^-\to\ell_2^-\ell_3^+\ell_4^-)=\frac{1}{1+\delta_{\ell_2\ell_4}}\frac{G_F^2 m_{\ell_1}^5}{3\cdot 2^{10}\pi^3}\times\\
\left\{
\abs{g_{LL}^{12,34}}^2+\abs{g_{LL}^{14,32}}^2+\abs{g_{RR}^{12,34}}^2+\abs{g_{RR}^{14,32}}^2+\abs{g_{LR}^{12,34}}^2+\abs{g_{LR}^{14,32}}^2\right.\\
\left.+\abs{g_{RL}^{12,34}}^2+\abs{g_{RL}^{14,32}}^2-\text{Re}\left[g_{LL}^{12,34}{g_{LL}^{14,32}}^\ast+g_{RR}^{12,34}{g_{RR}^{14,32}}^\ast\right]
\right\}\,.
\end{multline}
Experimental bounds on the corresponding branching ratios are collected in appendix \ref{AP:Input}.

\subsubsection{Neutral Meson mixings}
 The NP short distance tree level contribution to the meson-antimeson transition amplitude\footnote{$M$ is the hermitian part of the effective hamiltonian describing the evolution of the two-level, meson-antimeson, system; $M_{12}$ is the dispersive transition amplitude.} $M_{12}^{NP}$ is \cite{Lavoura}
\begin{multline}
M_{12}^{NP}=\\
\sum_{H=R,I}
\frac{f_M^2m_M}{96v^2m^2_H}\left(\left(1+\left(\frac{m_M}{m_{q_1}+m_{q_2}}\right)^2\right)C_1(H)-\left(1+11\left(\frac{m_M}{m_{q_1}+m_{q_2}}\right)^2\right)C_2(H)\right)
\end{multline}
where $C_1(R)=(N_{q_2q_1}^*+N_{q_1q_2})^2$, $C_2(R)=(N_{q_2q_1}^*-N_{q_1q_2})^2$,
$C_1(I)=-(N_{q_2q_1}^*-N_{q_1q_2})^2$ and $C_2(I)=-(N_{q_2q_1}^*+N_{q_1q_2})^2$.
$q_1$ and $q_2$ refer to the valence quarks of the corresponding meson and $N$ is $N_u$ or $N_d$ for up-type or down-type quarks (and thus mesons).
\begin{figure}[h]
\begin{center}
\includegraphics[width=0.325\textwidth]{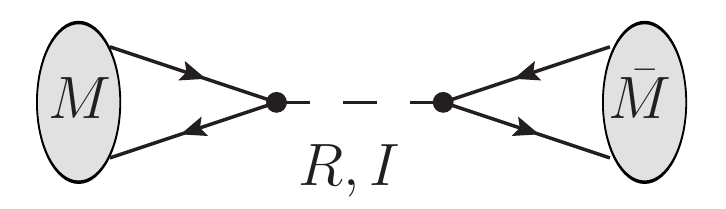}
\caption{Tree level $R,I$ mediated NP contributions to $M\to \bar M$.\label{fig:MtoMbar}}
\end{center}
\end{figure}
For both $B^0_d$--$\bar B^0_d$ and $B^0_s$--$\bar B^0_s$ systems, the mass differences $\Delta M_{B_d}$ and $\Delta M_{B_s}$ are, to a very good approximation (namely $M_{12}^{B_q}\gg \Gamma_{12}^{B_q}$ with $\Gamma_{12}^{B_q}$ the absorptive transition amplitude), 
\[
\Delta M_{B_d}=2\abs{M_{12}^{B_d}}\,,\qquad \Delta M_{B_s}=2\abs{M_{12}^{B_s}}\,.
\]
In addition, time dependent CP violating asymmetries in $B^0_d\to J/\Psi K_S$ and $B^0_s\to J/\Psi \Phi$ decays constrain the phase of $M_{12}^{B_d}$ and $M_{12}^{B_s}$, respectively. We incorporate neutral B meson mixing constraints through the quantities
\[
\Delta_d=\frac{M_{12}^{B_d}}{[M_{12}^{B_d}]_{\rm SM}}\,,\qquad \Delta_s=\frac{M_{12}^{B_s}}{[M_{12}^{B_s}]_{\rm SM}}\,,
\]
according to \cite{CKMfitter}.

In $K^0$--$\bar K^0$, both $M_{12}^K$ and $\Gamma_{12}^K$ are relevant for the mass difference and thus we require that the NP contribution to $M_{12}^K$ 
does not exceed the experimental value of $\Delta M_K$.  In addition we take into
account the CP violating observable $\epsilon_K$,
\[
|\epsilon_K| = \frac{\text{Im}(M_{12}^K)}{\sqrt{2}\Delta M_K}\,,
\]
where the new contribution cannot exceed 10\% of the experimental value.

For $D^0$--$\bar D^0$ long distance effects also prevent a direct connection between $M_{12}^D$ and $\Delta M_D$; as in $K^0$--$\bar K^0$, we then require  that the short distance NP contribution to $M_{12}^D$ does not give, alone, too large a contribution to
 $\Delta M_D$. In addition, it can be checked that the new contributions to CP violation in $D^0$--$\bar D^0$ are negligible. Since this is the only existing up-type neutral meson system, the constraints on flavour changing neutral couplings arising from neutral meson mixings are tighter for neutral couplings to down quarks than they are for up quarks. The values used in the analysis are collected in appendix \ref{AP:Input}.

\subsubsection{Rare decays $M^0\to\ell_1^+\ell_2^-$}
 Let us now consider mesons $M^0$ with valence quark composition $\bar q_2 q_1$
In BGL models, the tree level induced NP terms in the effective Lagrangian relevant for the rare decays $M^0\to\ell_1^+\ell_2^-$ are:
\begin{equation}
\mathcal L_{\rm eff}^{NP} = 
-\frac{2 G_F}{\sqrt{2}}\sum_{\chi_1,\chi_2=L,R}\ c_{\chi_1\chi_2}^{12,12} \big[\bar q_2\gamma_{\chi_1} q_1\big]\left[\bar\ell_2\gamma_{\chi_2} \ell_1\right]
\end{equation}
with
\begin{align*}
&c_{LL}^{ij,kl}=\frac{(N_q^\dagger)_{q_j q_i}(N_\ell^\dagger)_{\ell_l \ell_k}}{m_R^2}-\frac{(N_q^\dagger)_{q_j q_i}(N_\ell^\dagger)_{\ell_l \ell_k}}{m_I^2}\,,\ &
c_{RL}^{ij,kl}=\frac{(N_q)_{q_j q_i}(N_\ell^\dagger)_{\ell_l \ell_k}}{m_R^2}+\frac{(N_q)_{q_j q_i}(N_\ell^\dagger)_{\ell_l \ell_k}}{m_I^2}\,,\\
&c_{LR}^{ij,kl}=\frac{(N_q^\dagger)_{q_j q_i}(N_\ell)_{\ell_l \ell_k}}{m_R^2}+\frac{(N_q^\dagger)_{q_j q_i}(N_\ell)_{\ell_l \ell_k}}{m_I^2}\,,\ &
c_{RR}^{ij,kl}=\frac{(N_q)_{q_j q_i}(N_\ell)_{\ell_l \ell_k}}{m_R^2}-\frac{(N_q)_{q_j q_i}(N_\ell)_{\ell_l \ell_k}}{m_I^2}\,.
\end{align*}
Notice that for the lepton flavour violating modes $M^0\to\ell_1^+\ell_2^-$ with $\ell_1\neq\ell_2$, the SM contribution to the effective Lagrangian is absent, this is no longer
true in $\ell_2=\ell_1$ case.
\begin{figure}[h]
\begin{center}
\includegraphics[width=0.35\textwidth]{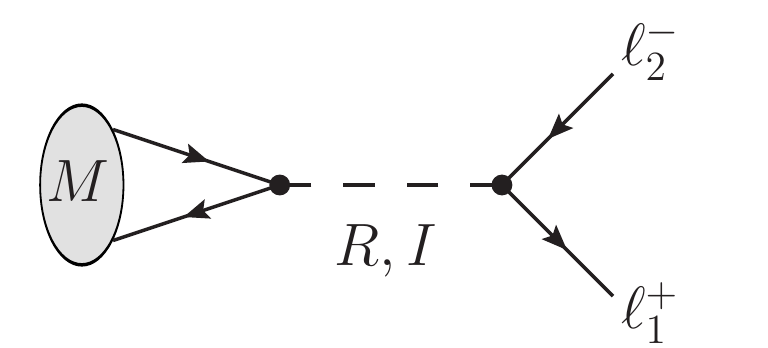}
\caption{Tree level $R,I$ mediated NP contributions to $M\to \ell_1^+\ell_2^-$.\label{fig:Mtoll2}}
\end{center}
\end{figure}

In the notation of appendix 7 of reference \cite{Crivellin:2013wna}, the Wilson coefficients read
\begin{align*}
&C_S^{q_2q_1} = -\frac{\sqrt{2}\pi^2}{G_FM_W^2}\left(c_{LR}^{12,12}+c_{LL}^{12,12}\right) \,, 
&C_P^{q_2q_1} = -\frac{\sqrt{2}\pi^2}{G_FM_W^2}\left(c_{LR}^{12,12}-c_{LL}^{12,12}\right)\,,\\
&C_S^{\prime\,q_2q_1} = -\frac{\sqrt{2}\pi^2}{G_FM_W^2}\left(c_{RR}^{12,12}+c_{RL}^{12,12}\right)\,,
&C_P^{\prime\,q_2q_1} = -\frac{\sqrt{2}\pi^2}{G_FM_W^2}\left(c_{RR}^{12,12}-c_{RL}^{12,12}\right)\,.
\end{align*}

The different modes and measurements used in the analysis are collected in appendix \ref{AP:Input}. It should be noted that while the previous type of short distance 
contributions dominate the rate for $B_s$ and $B_d$ decays, the situation is more involved in other cases. For example, for $K_L\to\mu^+\mu^-$ decays, the rate is dominated by the intermediate $\gamma\gamma$ state \cite{kaons} and NP is constrained through the bounds on the short distance SM+NP contributions.

\subsection{Loop level processes\label{SEC:ExpConst-sSEC:Loop}}
In the previous subsections we have listed observables useful to constrain the flavour changing couplings of the BGL models; their common characteristic is the possibility of having
NP contributions at tree level. In this subsection we address  two important rare decays where NP only contributes at loop level: $\ell_j\to\ell_i\gamma$ and $B\to X_s\gamma$.

\subsubsection{$\ell_j \to \ell_i \gamma$}
 Lepton flavour violating (LFV) processes like $\mu\to e\gamma$ or $\tau\to \mu\gamma$ are in general a source of severe constraints for models with FCNC, like the BGL models we are considering in this work. The reason, anticipated for $\ell_1\to \ell_2\bar\ell_3\ell_4$ decays, is that these processes are negligible in the SM (their amplitudes are proportional to $m_{\nu_k}^2/m_W^2\ll 1$), while in the BGL case we expect loop contributions from neutral Higgs  flavour changing couplings proportional to
$m_{\ell_k}^2/m_{R,I}^2$. Moreover, and contrary to other 2HDM, the charged Higgs can also be relevant here, as the non-unitarity of the matrices controlling the couplings $H^{-}\bar{\ell}_j\nu_k$ and $H^{+}\bar{\nu}_k\ell_i$ leads to contributions 
proportional to $ m_{\ell_j}m_{\ell_i}/m_{H^{\pm}}^2$ (which would otherwise cancel out when summing over all generations of neutrinos running in the loop).
\begin{figure}[h]
\begin{center}
\subfigure[$H^\pm$ mediated.\label{fig:a}]{\includegraphics[width=0.35\textwidth]{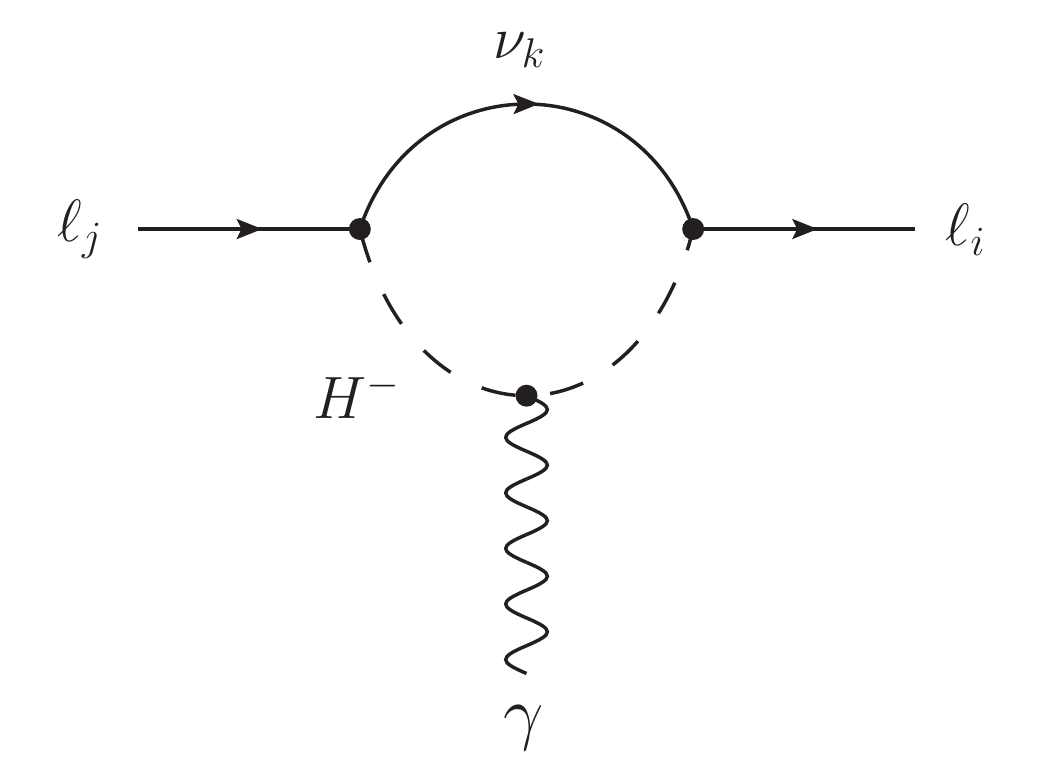}}\qquad
\subfigure[$R,I$ mediated.\label{fig:b}]{\includegraphics[width=0.35\textwidth]{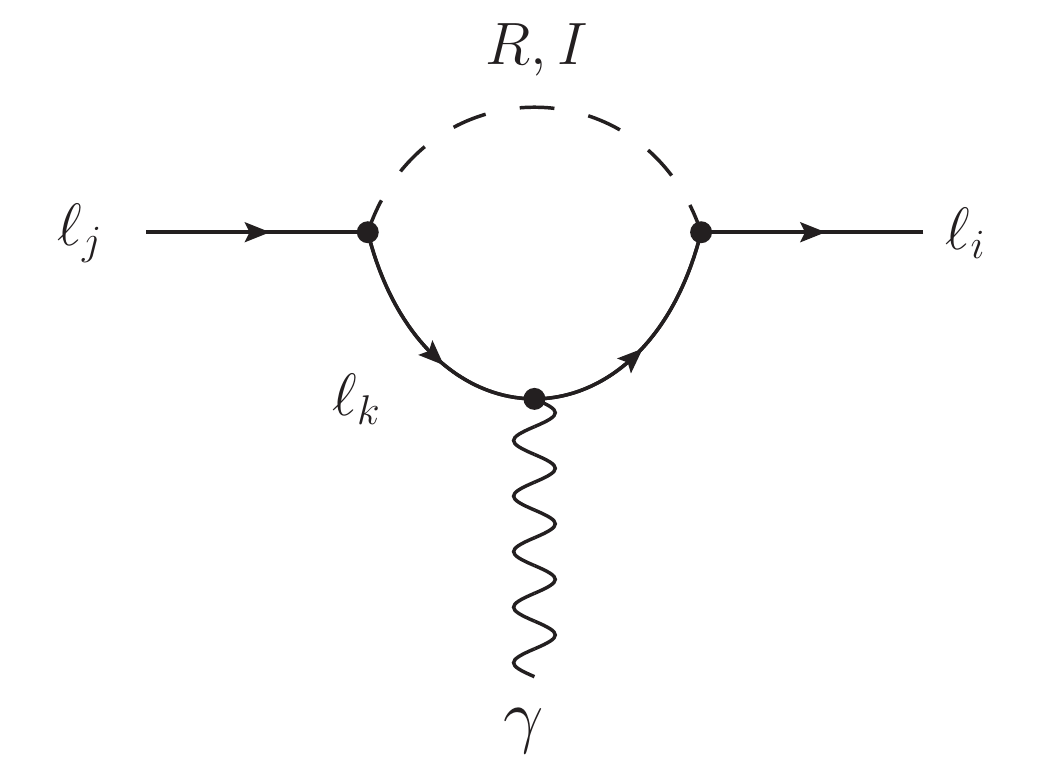}}
\caption{NP contributions to $\ell_j\to\ell_i\gamma$.\label{fig:ljtoli_g}}
\end{center}
\end{figure}
For on-shell photon and external fermions, the $\ell_j\to \ell_i\gamma$ amplitude is completely described by a dipole transition, see e.g. \cite{delAguila:2008zu},
\begin{equation}
i\mathcal{M}=ie\left[\mathcal{A}_R \gamma_R+ \mathcal{A}_L\gamma_L\right]\sigma^{\mu\nu}q_{\nu}\epsilon_{\mu},
\end{equation}
with $q^{\mu}$ the incoming photon momentum. The corresponding decay width is
\begin{equation}
\Gamma(\ell_j \to \ell_i\gamma)=\frac{\alpha m_{\ell_j}^5 G_F^2}{128\pi^4}\left[\left|\mathcal{A}_L\right|^2+\left|\mathcal{A}_R\right|^2\right]\,.
\end{equation}
Up to terms of $\mathcal{O}(m_{\ell_i}/m_{\ell_j})$ -- note that
$N_\ell^{ik}$ is proportional to $m_{\ell_k}$ --, the coefficients $\mathcal{A}_R$ and $\mathcal{A}_L$ are given by
\begin{multline}
\mathcal{A}_R=
\sum_{k}\left\{\frac{1}{12 m_{R}^2}N_\ell^{ik}N_\ell^{jk\ast}
-\frac{1}{2m_{R}^2}N_\ell^{ik}N_\ell^{kj}\frac{m_{\ell_k}}{m_{\ell_j}}
\left[\frac{3}{2}+\ln\left(\frac{m_{\ell_k}^2}{m_{R}^2}\right)\right]\right.\\
+\left.\frac{1}{12 m_I^2}N_\ell^{ik}N_\ell^{jk\ast}
+\frac{1}{2m_{I}^2}N_\ell^{ik}N_\ell^{k j}\frac{m_{\ell_k}}{m_{\ell_j}}\left[\frac{3}{2}+\ln\left(\frac{m_{\ell_k}^2}{m_{I}^2}\right)\right]\right\},
\end{multline}
\begin{multline}
\mathcal{A}_L=
\sum_{k}\left\{-\frac{1}{12m_{H^{\pm}}^2}(N_\ell^{\dagger}U)^{i k}(N_{\ell}^{\dagger}U)^{j k\ast}
+\frac{1}{12m_{R}^2}N_{\ell}^{ki\ast}N_{\ell}^{kj}
+\frac{1}{12m_{I}^2}N_{\ell}^{k i\ast}N_{\ell}^{k j}\right\},
\end{multline}
where we have neglected contributions proportional to the neutrino masses $m_{\nu_k}\approx 0$ as well as subleading terms in $m_{\ell_k}^2/m_{R,I}^2$.  

In some cases, two-loop contributions for $\ell_{j}\to \ell_{i}\gamma$ can dominate over the one-loop ones \cite{Bjorken:1977vt,Chang:1993kw}. This is related to the fact that, due to the required chirality flip, we need three mass insertions at one loop level. However, there are two-loop contributions with only one chirality flip in the $\ell_j-\ell_i$ fermion line. Therefore, in some cases they can compensate the extra loop factor
by avoiding two small Yukawa couplings. We can roughly estimate the two-loop contribution as 
\begin{eqnarray}
\Gamma(\ell_j\to \ell_i\gamma)_{\mathrm{2-loop}}\approx \frac{\alpha m_{\ell_j}^5 G_F^2}{128 \pi^4}\left(\frac{\alpha}{\pi}\right)^2[\left|\mathcal{C}\right|^2+\left|\mathcal{D}\right|^2],
\end{eqnarray}
where
\begin{equation}
\mathcal{C}=\frac{2}{m_{R}^2}
(N_{\ell})_{ij}(N_{u})_{tt}\frac{m_t}{m_{\ell_j}}\ln^2\left(\frac{m_t^2}{m_{R}^2}\right)\quad
\text{and}\quad
\mathcal{D}=\frac{2}{m_{I}^2}
(N_{\ell})_{ij}(N_{u})_{tt}\frac{m_t}{m_{\ell_j}}\ln^2\left(\frac{m_t^2}{m_{I}^2}\right).
\end{equation}

\subsubsection{$\bar B\to X_s\gamma$}
 The other important rare decay, now in the quark sector, is $\bar B\to X_s\gamma$, induced by the quark level transition $b\to s\gamma$.
Similarly to the LFV processes $\ell_j\to\ell_i\gamma$ considered before, NP contributions due to the exchange of both neutral and charged Higgs are present. Although the contributions coming from the latter case are naively expected to be dominant, due to the relative enhancement coming from  the top mass insertion -- i.e. proportional to
$ m_{t}^2/m_{H^{\pm}}^2$ versus $ m_{b}^2/m_{R,I}^2$ --, we cannot neglect diagrams with FCNC because this effect can be compensated by $\tan\beta$ enhancements. The effective Hamiltonian describing this transition is
\begin{equation}
\mathcal{H}_{\rm eff}(b\to s\gamma)=-\frac{4 G_F}{\sqrt{2}} V_{tb}V_{ts}^{\ast}\left[C_7(\mu_b)\mathcal{O}_7+C_7^{\prime}(\mu_b)\mathcal{O}_7^{\prime}+C_8(\mu_b)\mathcal{O}_8+C_8^{\prime}(\mu_b)\mathcal{O}_8^{\prime}\right],
\end{equation}
with new effective operators $\mathcal{O}_{7}^{\prime}$ and $\mathcal{O}_8^{\prime}$, which are absent in the SM besides terms $\mathcal{O}(m_s/m_b)$. $C_{7,8}(\mu)$ and $C_{7,8}^{\prime}(\mu)$ are the Wilson coefficients of the dipole operators
\begin{eqnarray}
\mathcal{O}_7&=&\frac{e}{16\pi^2}m_b \bar{s}_{L,\alpha}\sigma^{\mu\nu}b_{R,\alpha}F_{\mu\nu},\qquad	\mathcal{O}_8=\frac{g_s}{16\pi^2}m_b \bar{s}_{L,\alpha}\left(\frac{\lambda^a}{2}\right)_{\alpha\beta}\sigma^{\mu\nu}b_{R,\beta}G_{\mu\nu}^a,\\
\mathcal{O}_7^{\prime}&=&\frac{e}{16\pi^2}m_b \bar{s}_{R,\alpha}\sigma^{\mu\nu}b_{L,\alpha}F_{\mu\nu},\qquad  \mathcal{O}_8^{\prime}=\frac{g_s}{16\pi^2}m_b \bar{s}_{R,\alpha}\left(\frac{\lambda^a}{2}\right)_{\alpha\beta}\sigma^{\mu\nu}b_{L,\beta}G_{\mu\nu}^a,\qquad~
\end{eqnarray}
evaluated at the scale $\mu_b=\mathcal{O}(m_b)$, with $F_{\mu\nu}$ and $G_{\mu\nu}^a$ 
denoting the electromagnetic and gluon field strength tensors, and $\lambda^{a}$, $a=1,
\ldots,8$, standing for the Gell-Mann matrices. 
\begin{figure}[h]
\begin{center}
\subfigure[$H^\pm$ mediated.\label{fig:1}]{\includegraphics[width=0.3\textwidth]{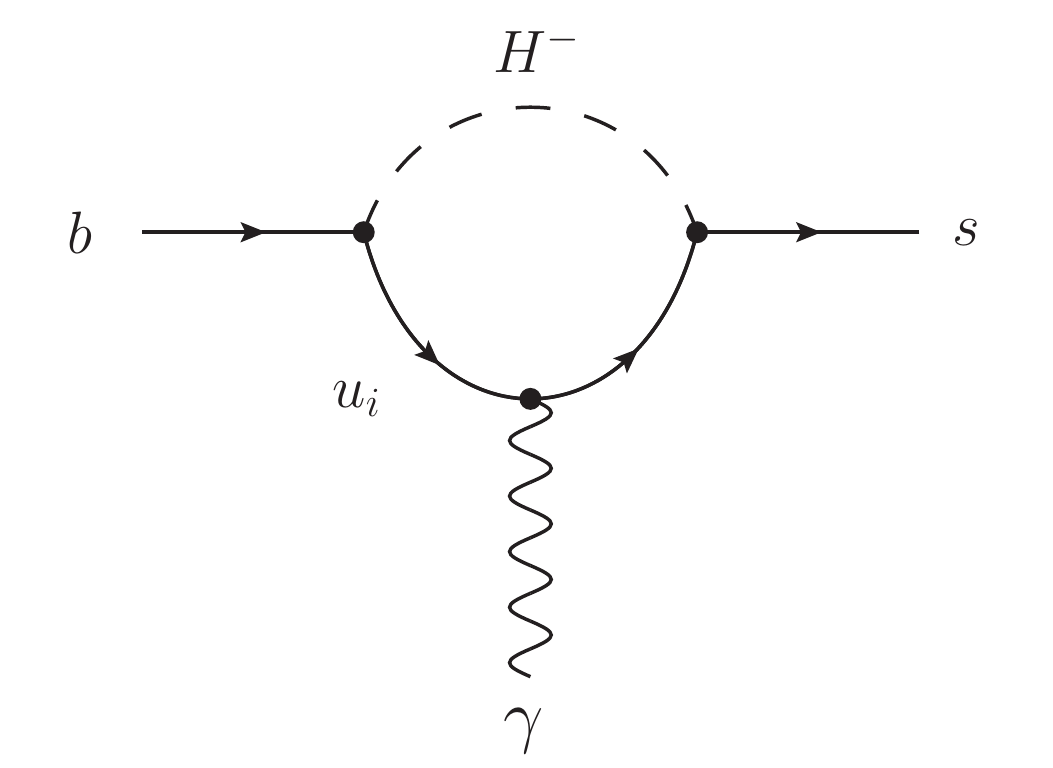}
\quad\includegraphics[width=0.3\textwidth]{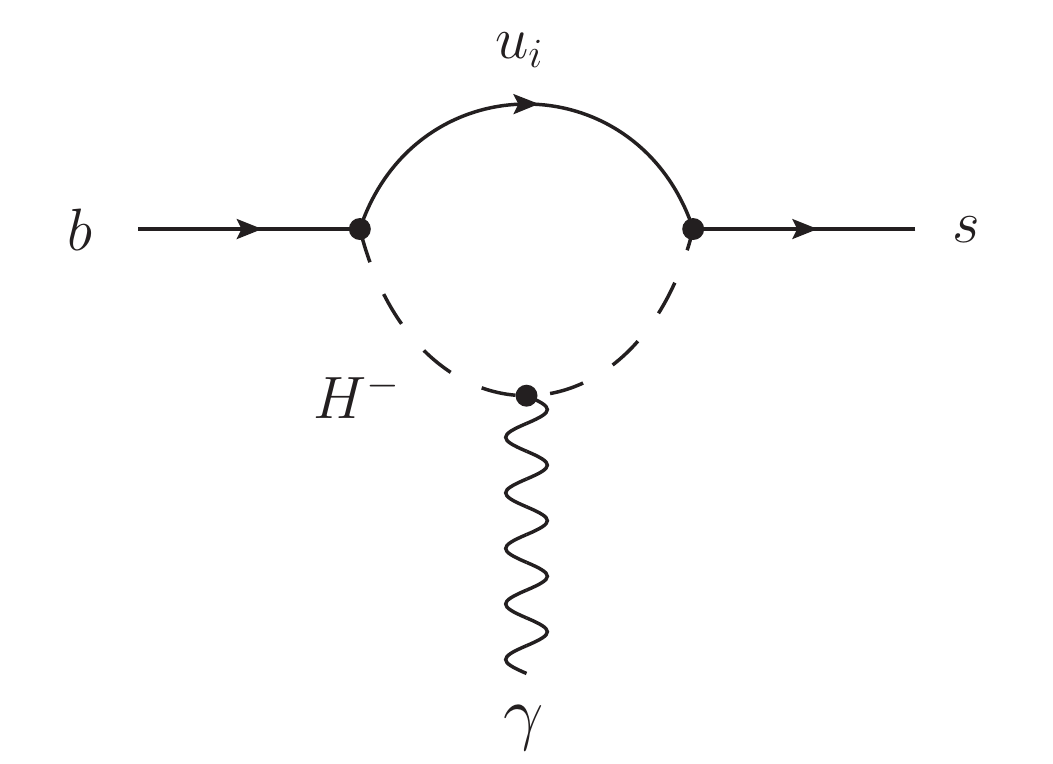}}\quad
\subfigure[$R,I$ mediated.\label{fig:2}]{\includegraphics[width=0.3\textwidth]{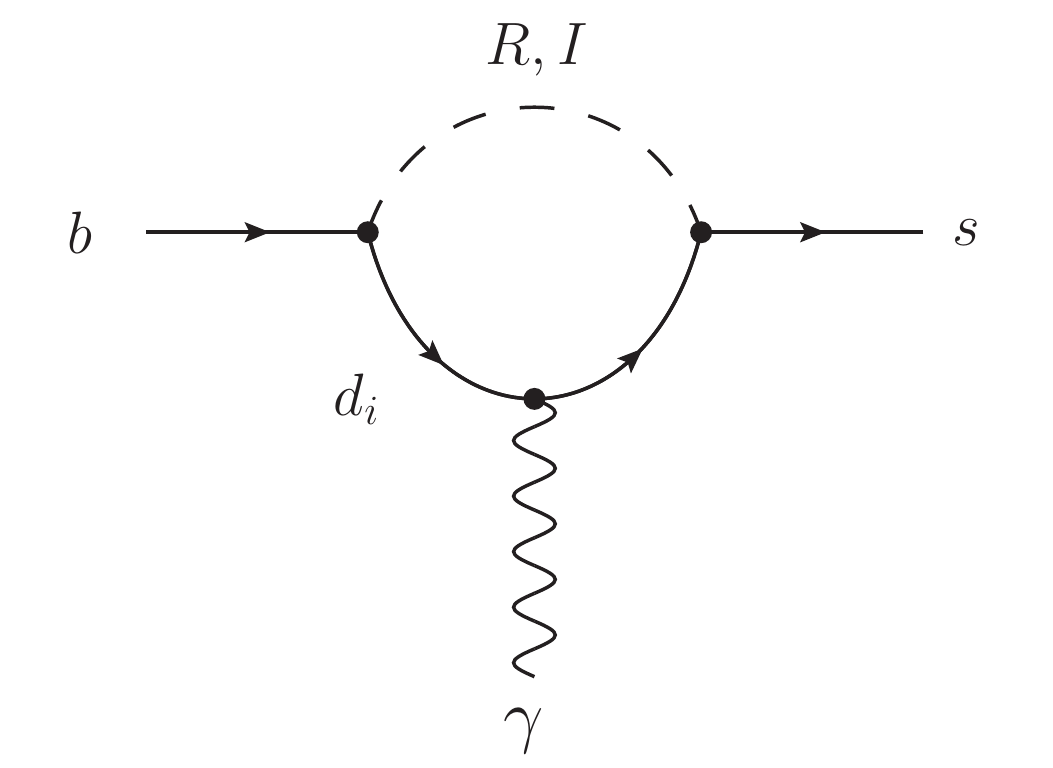}}
\caption{NP contributions to $b\to s\gamma$.\label{fig:btosg}}
\end{center}
\end{figure}

We then constrain the BGL contribution to $b\to s\gamma$ using the master formula \cite{Misiak:2006ab, Buras:2011zb, Blanke:2011ry, Blanke:2012tv}
\begin{equation}
\mathrm{Br}\left(\bar{B}\to X_s\gamma\right)=\mathrm{Br}_{\rm SM}+0.00247\left[|\Delta C_7(\mu_b)|^2+|\Delta C_7^{\prime}(\mu_b)|^2-0.706\mathrm{Re} \left(\Delta C_7(\mu_b)\right)\right],\label{eq:bsgamma}
\end{equation}
where $\mathrm{Br}_{\rm SM}=\mathrm{Br}(\bar{B}\to X_s\gamma)_{\rm SM}=(3.15\pm 0.23)\times 10^{-4}$ is the SM prediction at NNLO \cite{Gambino:2001ew, Misiak:2006zs, Misiak:2006ab} and we have split the SM and the NP contributions to the relevant Wilson coefficients
\begin{equation}
C_{7}^{(\prime)}(\mu)=C_{7,\mathrm{SM}}^{(\prime)}(\mu)+\Delta C_7^{(\prime)}(\mu), \qquad 	C_{8}^{(\prime)}(\mu)=C_{8,\mathrm{SM}}^{(\prime)}(\mu)+\Delta C_8^{(\prime)}(\mu).
\end{equation}
The value obtained from equation (\ref{eq:bsgamma}) has to be compared with the experimental measurement \cite{Asner:2010qj}
\begin{equation}
\mathrm{Br}(B\to X_s \gamma)_{\rm exp}= \left(3.55 \pm 0.27\right) \times 10^{-4}.
\end{equation}
The Wilson coefficients  $\Delta C_{7,8}^{}(\mu)$ and $\Delta C_{7,8}^{\prime}(\mu)$ are computed at the high energy scale $\tilde{\mu}=\mathcal{O}(m_{H^{\pm}})\sim\mathcal{O}(m_{R,I})$ at one-loop in perturbation theory, and then run down to $\mu_b$ using RGE \cite{Grinstein:1990tj,Buras:1993xp}:
\begin{equation}
\Delta C_7^{(\prime)}(\mu_b)\approx \eta^{\frac{16}{23}}\Delta C_7^{(\prime)}(\tilde{\mu})+\frac{8}{3}\left(\eta^{\frac{14}{13}}-\eta^{\frac{16}{23}}\right)\Delta C_8^{(\prime)}(\tilde{\mu}),
\end{equation}
where $\eta=\alpha_s(\tilde{\mu})/\alpha_s(\mu_b)$. FCNC might also affect the running of these Wilson coefficients through new operators which are not present in the SM, similarly to what happens in the case of flavour changing neutral gauge bosons \cite{Buras:2011zb}. However, the impact of this effect is expected to be subleading, and its study is well beyond the scope of this paper. 

The relevant Wilson coefficients read
\begin{multline}
\Delta C_7(\tilde{\mu})=\frac{1}{2}\frac{1}{V_{ts}^\ast V_{tb}}
\sum_k\left\{\frac{1}{m_{H^{\pm}}^2}(V^{\dagger}N_{u})_{sk}
\left((V^{\dagger}N_{u})^{\ast}_{bk}A^{(2)}_H(x_{H^{\pm}}^k)
+(N^{\dagger}_d V^{\dagger})^{\ast}_{bk}\frac{m_{u_k}}{m_{b}}A_H^{(3)}(x_{H^{\pm}}^k)\right)\right.\\
-(N_d)_{sk}(N_d)^{\ast}_{bk}\left(\frac{Q_{d}}{m_{R}^2}A_H^{(0)}(y_R^{k})+\frac{Q_{d}}{m_{I}^2}A^{(0)}_H(y_I^{k})\right)\\
-\left.(N_d)_{sk}(N_d)_{kb}\frac{m_{d_k}}{m_{b}}\left(\frac{Q_{d}}{m_{R}^2}A^{(1)}_H(y_R^k)-\frac{Q_{d}}{m_{I}^2}A^{(1)}_H(y_I^k)\right)\right\},
\end{multline}
\begin{multline}
\Delta C_7^{\prime}(\tilde{\mu})=\frac{1}{2}\frac{1}{V_{ts}^\ast V_{tb}}
\sum_k\left\{\frac{1}{m_{H^{\pm}}^2}(N_d^{\dagger}V^{\dagger})_{sk}(N_d^{\dagger}V^{\dagger})^{\ast}_{bk}A_{H}^{(2)}(x_{H^{\pm}}^k)\right.\\
-(N_d)^{\ast}_{ks}(N_d)_{kb}\left.\left(\frac{Q_d}{m_{R}^2}A_H^{(0)}(y_R^{k})+\frac{Q_d}{m_{I}^2}A^{(0)}_H(y_I^{k})\right)\right\},
\end{multline}
\begin{multline}
\Delta C_8(\tilde{\mu})=\frac{1}{2}\frac{1}{V_{ts}^\ast V_{tb}}
\sum_k\left\{2\frac{1}{m_{H^{\pm}}^2}(V^{\dagger}N_{u})^{sk}\left(-(V^{\dagger}N_{u})^{\ast}_{bk}A^{(0)}_H(x_{H^{\pm}}^k)+(N^{\dagger}_d V^{\dagger})^{\ast}_{bk}\frac{m_{u_k}}{m_{b}}A^{(1)}_H(x_{H^{\pm}}^k)\right)\right.\\
-(N_d)_{sk}(N_d)^{\ast}_{bk}\left(\frac{1}{m_{R}^2}A_H^{(0)}(y_R^{k})+\frac{1}{m_{I}^2}A^{(0)}_H(y_I^{k})\right)\\
-\left.(N_d)_{sk}(N_d)_{kb}\frac{m_{d_k}}{m_{b}}\left(\frac{1}{m_{R}^2}A_H^{(1)}(y_R^{k})-\frac{1}{m_{I}^2}A^{(1)}_H(y_I^{k})\right)\right\},
\end{multline}
\begin{multline}
\Delta C_8^{\prime}(\tilde{\mu})=\frac{1}{2}\frac{1}{V_{ts}^\ast V_{tb}}
\sum_k\left\{-2\frac{1}{m_{H^{\pm}}^2}(N_d^{\dagger}V^{\dagger})_{sk}(N_d^{\dagger}V^{\dagger})^{\ast}_{bk}A_{H}^{(0)}(x_{H^{\pm}}^k)\right.\\
-(N_d)^{\ast}_{ks} (N_d)_{kb}\left.\left(\frac{1}{m_R^2}A_{H}^{(0)}(y_R^k)+\frac{1}{m_I^2}A_{H}^{(0)}(y_I^k)\right)\right\},
\end{multline}
where $Q_d=-1/3$ and $x_{H^{\pm}}^k=m_{u_k}^2/m_{H^{\pm}}^2$, $y_{R,I}^k=m_{d_k}^2/m_{R,I}^2$. The loop functions $A_H^{(i)}$ are:

\begin{align*}
&A_{H}^{(0)}(x)=\frac{2+3x-6x^2+x^3+6x\ln x}{24(1-x)^4},\quad  A_H^{(2)}(x)=\frac{-7+5x+8x^2}{36(1-x)^3}+\frac{x(-2+3x)\ln x}{6(1-x)^4},\\
&A_{H}^{(1)}(x)=\frac{-3+4x-x^2-2\ln x}{4(1-x)^3},\quad A_H^{(3)}(x)=\frac{-3+8x-5 x^2-(4-6 x)\ln x}{6 (1-x)^3}.
\end{align*}

\subsubsection{Electric dipole moments and anomalous magnetic moments}

 NP induced one loop contributions to the electric dipole moments (EDM) of leptons and quarks are absent in BGL models. In \cite{Botella:2012ab} it has been shown that the weak basis invariant relevant for the quark EDMs does not develop an imaginary part. Two loop diagrams including strong corrections to the one loop invariants do not change the situation, therefore it is also trivial that in BGL models there is no contribution to the Weinberg operator. In fact, if we take the general parametrizations of the Higgs couplings to fermions used in \cite{Jung:2013hka} it turns out that all the parameters whose imaginary part contribute to the EDMs become real in the BGL models studied here. That is, even at two loops, EDMs are zero in BGL models. For the anomalous magnetic moments, we checked that the NP induced one loop contributions appearing in BGL models are too small to have significant impact on the results -- once other constraints are used --, in agreement with \cite{Crivellin:2013wna}.

\subsubsection{Precision Electroweak Data}
 The previous subsections have covered representative flavour related low energy processes that are able to constrain the masses of the new scalar together with $\tan\beta$.
Electroweak precision data also play an important  role. The observables included in the analysis for that purpose are the $Z\bar bb$ effective vertex and the oblique parameters $S$, $T$ and $U$.

For the $Z\bar bb$ vertex probed at LEP, BGL models introduce new contributions mediated by the charged and by the neutral scalars. The effects mediated by $H^\pm$  are typically the most relevant ones, see e.g. \cite{HernandezSanchez:2012eg}.  In our case, similarly to what happens in $b\to s\gamma$, neutral contributions can also be relevant but, as a first estimate, we just consider the charged ones \cite{Pich:2010}
\begin{equation}
F_{Zb\bar b}=\frac{|C^{tb}|-0.72}{m_H^\pm}< 0.0024\text{ \GeV}^{-1}\,,
\end{equation}
where once again $C^{tb}=-1/\tan\beta$ for BGL models of quark types $t$ and $b$, and $C^{tb}=\tan\beta$ otherwise.

For the oblique parameters, as discussed in \cite{O'Neil:2009nr}, the contributions to $S$ and $U$ in 2HDM tend to be small. This is not the case for the $T$ parameter which receives corrections that can be sizable. In BGL models, the NP contribution $\Delta T$ to $T=T_{{\rm SM}}+\Delta T$  \cite{Grimus:2007if, Grimus:2008nb} is
\begin{equation}
\Delta T=\frac{1}{16 \pi m_W^2s_W^2}\left\{F(m_{H^{\pm}}^2,m_{R}^2)-F(m_I^2,m_{R}^2)+F(m_{H^{\pm}}^2,m_{I}^2)\right\}
\label{drho2}
\end{equation}
with
\[
F(x,y)=\frac{x+y}{2}-\frac{xy}{x-y}\ln\frac{x}{y}\,,
\]
so that $F(x,x)= 0$, while for $\Delta S$
\begin{equation}
\Delta S=\frac{1}{24\pi}\left\{\left(2s_W^2-1\right)^2G(m_{H^{\pm}}^2,m_{H^{\pm}}^2,m_Z^2)+G(m_{R}^2,m_{I}^2,m_Z^2)+\ln\left[\frac{m_{R}^2m_{I}^2}{m_{H^{\pm}}^4}\right]\right\},\qquad 
\end{equation}
where
\begin{multline}
G(x,y,z)=-\frac{16}{3}+5\frac{x+y}{z}-2\frac{(x-y)^2}{z^2}+\frac{r}{z^3}f(t,r)\\
    +\frac{3}{z}\left[\frac{x^2+y^2}{x-y}-\frac{x^2-y^2}{z}+\frac{(x-y)^3}{3z^2}\right]\ln\frac{x}{y},
\end{multline}
with $r=z^2-2z(x+y)+(x-y)^2$, $t=x+y-z$ and
\begin{equation}
f(t,r)=
\begin{cases}
\sqrt{r}\ln\left|\frac{t-\sqrt{r}}{t+\sqrt{r}}\right|&r>0,\\
2 \sqrt{-r}\arctan\frac{\sqrt{-r}}{t}&r<0.
\end{cases}
\end{equation}

%
%

\clearpage
\section{Results\label{SEC:Results}}
\begin{figure}[!htb]
\centering
\includegraphics[width=0.45\textwidth]{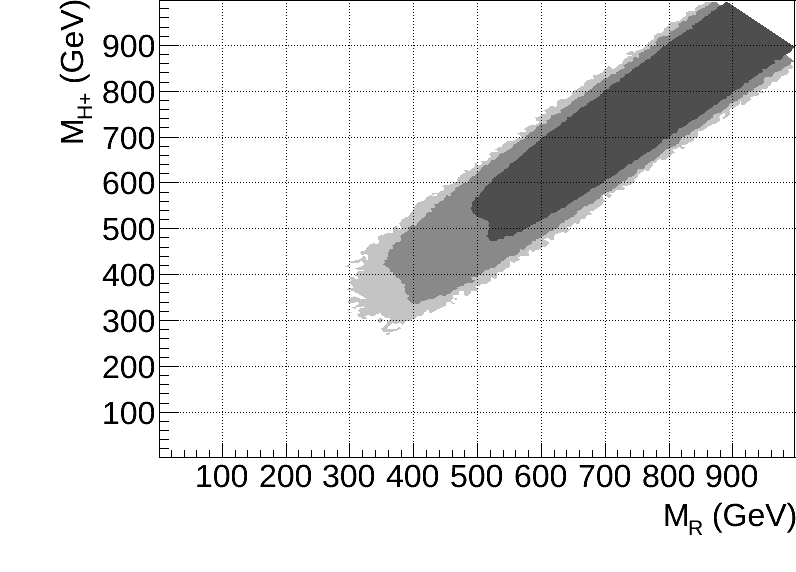}
\includegraphics[width=0.45\textwidth]{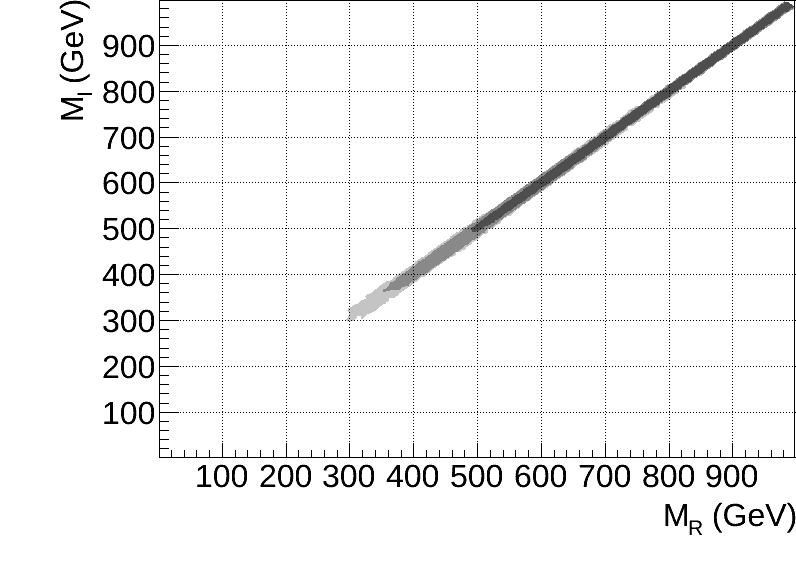}
\caption{Effect of the oblique parameters constraints in model $(t,\tau)$.}
\label{fig:onea}
\end{figure}

In the previous section we have presented a large set of relevant observables that can constrain the different BGL models, excluding regions of the parameter space $\{\tan\beta, m_{H^\pm},$ $m_R, m_I\}$ where the NP contributions are not compatible with the available experimental information. Following the methodology described in appendix \ref{AP:Analysis}, we apply those constraints to each one of the 36 BGL models: the main aim of this general study is to understand where could the masses of the new scalars lie and how does this depend on $\tan\beta$. However, before addressing the main results for the complete set of BGL models, an important aspect has to be settled: since we have three different scalars, we should in principle obtain allowed regions in the $\{\tan\beta, m_{H^\pm}, m_R, m_I\}$ parameter space, and then project them to the different subspaces for each BGL model, e.g. $m_{H^\pm}$ vs. $\tan\beta$, $m_{R}$ vs. $\tan\beta$, etc. The oblique parameters, in particular $\Delta T$, help us to simplify the picture. For degenerate $H^\pm$, $R$ and $I$, according to Eq.~(\ref{drho2}), $\Delta T=0$; in general, for almost degenerate $H^\pm$, $R$ and $I$, the oblique parameters are in agreement with experimental data\footnote{$\Delta T$ alone is not sufficient; considering only $\Delta T$, for $m_{H^\pm}=m_I$, $m_R$ would be free to vary but $\Delta S$ prevents it.  Analogously, for $m_{H^\pm}=m_R$, $\Delta T=0$ irrespective of
$m_I$.
In addition, in the experimental constraint, $\Delta T$ and $\Delta S$ are correlated.}. This is explored and illustrated in figure \ref{fig:onea} for one particular model: $m_R$ vs. $m_{H^\pm}$ and $m_R$ vs. $m_I$ allowed regions are displayed when the oblique parameters constraints are used.  Therefore, even though we treated all three scalar masses 
independently and on equal basis, we only present results in terms of $m_{H^\pm}$ for simplicity.

\begin{figure}[!htb]
\centering
\includegraphics[width=\textwidth]{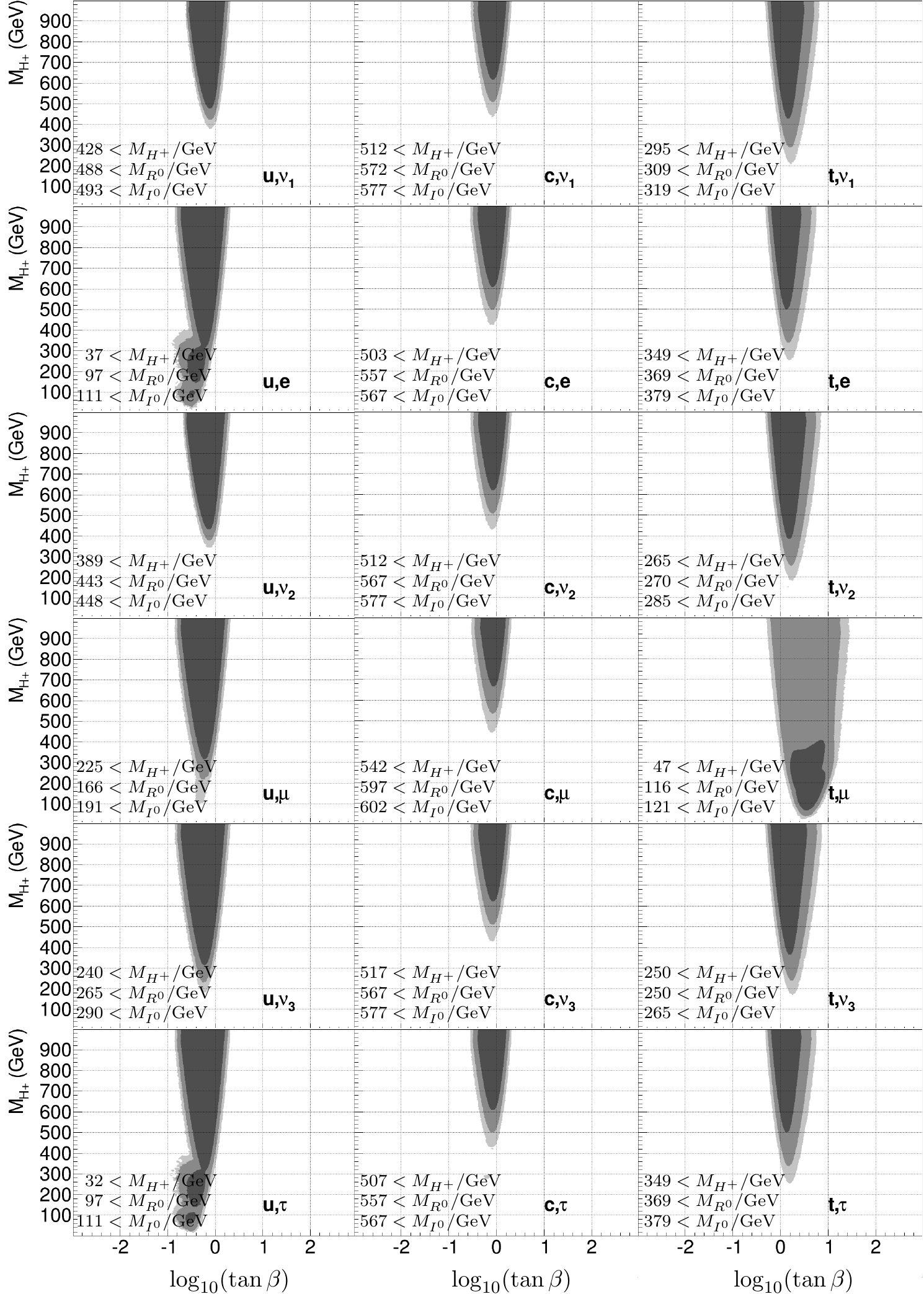}
\caption{Allowed 68\% (black), 95\% (gray) and 99\% (light gray) CL regions in $m_{H^\pm}$ vs. $\tan\beta$ for BGL models of types $(u_i,\nu_j)$ and $(u_i,\ell_j)$, i.e. for models with FCNC in the down quark sector and in the charged lepton or neutrino sector (respectively). Lower mass values corresponding to 95\% CL regions are shown in each case.\label{FIG:Results01}}
\end{figure}

\begin{figure}[!htb]
\centering
\includegraphics[width=\textwidth]{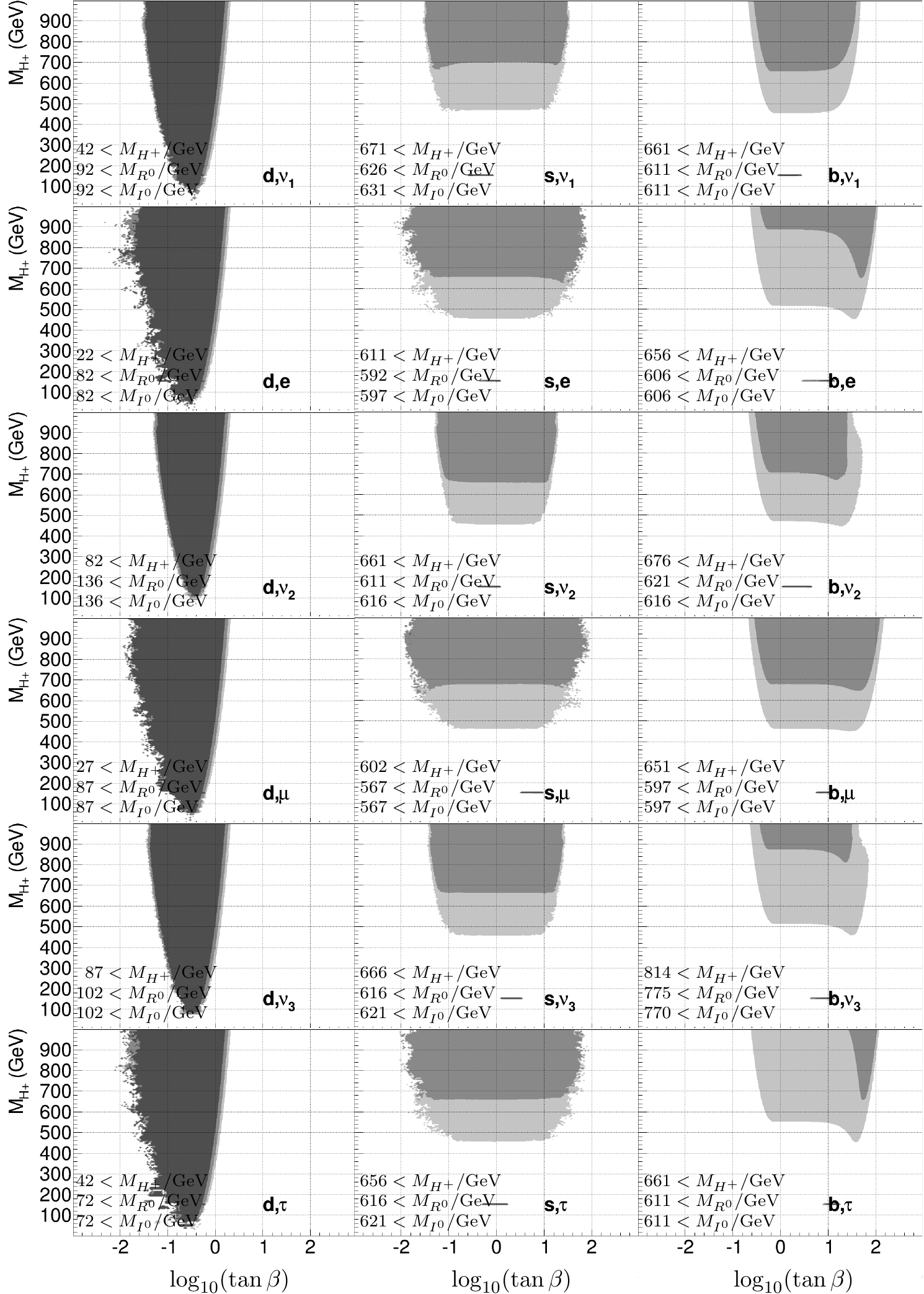}
\caption{Allowed 68\% (black), 95\% (gray) and 99\% (light gray) CL regions in $m_{H^\pm}$ vs. $\tan\beta$ for BGL models of types $(d_i,\nu_j)$ and $(u_i,\ell_j)$, i.e. for models with FCNC in the up quark sector and in the charged lepton or neutrino sector (respectively). Lower mass values corresponding to 95\% CL regions are shown in each case.\label{FIG:Results02}}
\end{figure}

In figures \ref{FIG:Results01} and \ref{FIG:Results02} we present the allowed regions -- corresponding to 68\%, 95\% and 99\% confidence levels (CL) -- in the $(m_{H^\pm},\tan\beta)$ plane for the 36 different BGL models. They deserve several comments.

\begin{itemize}
\item The experimental bounds for FCNC in the up sector are more relaxed than for the down sector, but the models with tree level FCNC in the up sector are not less constrained than the ones with tree level FCNC in the down sector, due to the $b\to s\gamma$ constraints on the charged Higgs mass. 

\item It should be emphasized that among the BGL models, the ones of types $t$ and $b$ guarantee a stronger suppression of the FCNC due to the hierarchical nature of the CKM matrix, so one would expect them to be less constrained. However, $b\to s\gamma$ frustrates this expectation. In fact, the models of type $d$ are less constrained than the $s$ and $b$ ones, while for up type models there is no clear trend.

\item Notice that some models allow for masses below the constraint $m_{H^+}>380$ GeV that $b\to s\gamma$ alone imposes on type II 2HDMs \cite{Hermann:2012fc}; this is due to the different $\tan\beta$ dependence of the contributions mediated by the charged scalar, which change from model to model. Neutral scalars play a very secondary role.

\item For the leptonic part, since the experimental bounds on tree level FCNC in the neutrino sector are irrelevant -- due to the smallness of neutrino masses --, $e$, $\mu$ and $\tau$ models are typically less constrained than $\nu_i$ models. This can be seen in figure \ref{FIG:Results01}, whereas in figure \ref{FIG:Results02} differences are minute, signifying then that leptonic constraints are secondary once other constraints are imposed.

\item Lower bounds on the scalar masses lie in between 100 and 400 GeV for many models, which put them within range of direct searches at the LHC. Nevertheless some exceptions deserve attention: for models of types $s$ and $b$, the lightest masses are instead in the 500-700 GeV range. Notice in addition that in models of types $s$ and $b$ the allowed values of $\tan\beta$ span a wider range than in the rest of models.

\item One aspect that is interesting on its own but would require specific attention beyond the scope of the present work, is the following: in many models isolated allowed regions for light masses appear. That is, for the considered set of observables, the scalar masses and $\tan\beta$ can still be tuned to agree with experimental data within these reduced regions. Higher order contributions than the ones used in section \ref{SEC:ExpConst}, additional observables and direct searches may then be used to further constrain these parameter regions.

\item As a final comment it should be noticed that some of the $t$ type models, the ones that correspond to the MFV framework as defined 
in \cite{Buras:2000dm} or \cite{D'Ambrosio:2002ex}, can be very promising. However
this is not a unique feature of these implementations since, as can be seen
from our figures, there are several others that allow for light scalars.


\end{itemize}

\clearpage

\clearpage
\clearpage

\section{Conclusions\label{sec:concl}}
We have presented phenomenological constraints on a class of models (BGL models) with two
Higgs doublets, where FCNC arise at tree level, but are naturally suppressed by $V_{\mathrm{CKM}}$
matrix elements. This is one of the remarkable features of BGL models which results from the
introduction, at Lagrangian level, of a discrete symmetry which constrains the
Yukawa couplings to have a special form. This symmetry can be implemented in the quark
sector in six different ways, and the same applies to the leptonic sector, leading altogether
to thirty six different realizations of the BGL models. The level of natural suppression
of FCNC is different in each of these realizations of BGL models and this obviously
leads to different constraints on the physical scalar masses allowed by experiment. 
Another interesting feature of BGL models is the fact that they have no other
flavour parameters, apart from CKM and PMNS matrix
elements. We study the allowed regions in the parameter space $\tan\beta$, 
$m_{H^+}$, $m_R$,  $m_I$ and then we project, for each BGL model, these regions into
subspaces relating pairs of the above parameters.  
Our results clearly show that this class of models
allow for new physical scalars beyond the standard Higgs boson, with masses which are
reachable, for example, at the next round of experiments at LHC.

For a long time, there was the belief that the only experimentally viable 2HDM extensions
of the SM were those where one has Natural Flavour Conservation in the Higgs
sector. BGL-type models provide an interesting alternative to NFC and the fact that they
allow for new scalars with masses within experimental reach, is specially exciting. 
In the BGL framework the number of additional free parameters introduced by
extending the scalar sector to two Higgs doublets is limited by the imposed discrete
symmetry. 

After this work was sent to the arXiv a paper was also submitted
to the arXiv  \cite{Bhattacharyya:2014nja}
analysing  one of the BGL scenarios discussed here, including
in addition the decay signatures of the new scalars. This paper agrees
with our conclusion concerning the feasibility of a light charged Higgs boson.

\clearpage

\acknowledgments
The work of FJB and MN was partially supported by Spanish MINECO under grant 
FPA2011-23596 and by \emph{Generalitat Valenciana} under grant GVPROMETEO 2010-056.
The work of GCB, LP and MNR was partially supported by
Funda\c c\~ ao  para a Ci\^ encia e a Tecnologia (FCT, Portugal) 
through the projects CERN/FP/123580/2011, PTDC/FIS-NUC/0548/2012, EXPL/FIS-NUC/0460/2013 
and CFTP-FCT Unit 777\\ (PEst-OE/FIS/UI0777/2013)
which are partially funded through POCTI (FEDER). 
The work of AC was supported by the Swiss National Science Foundation under contract 200021-143781.
MN is presently supported by a postdoctoral fellowship of the project CERN/FP/123580/2011.
LP is supported by FCT under contract SFRH/BD/70688/2010. 
The authors thank IFIC, Universidade de Valencia- CSIC, 
CFTP, Instituto Superior Tecnico, Universidade de Lisboa as well as ETH Zurich for hospitality
during their visits for collaboration work.
Part of the work done by MNR and GCB took place at CERN. 
The authors thank Luis Lavoura for useful conversations and the referee for constructive comments.

\appendix

\section{Analysis details\label{AP:Analysis}}
In figures \ref{fig:onea}, \ref{FIG:Results01} and \ref{FIG:Results02} we have presented 68\%, 95\% and 99\% CL allowed regions in parameter space. To wit, we represent regions where the specific BGL model is able to fit the imposed experimental information at least as well as the corresponding goodness levels. Some comments are in order. This procedure corresponds to the profile likelihood method \cite{asymptotic}. In brief, for a model with parameters $\vec p$, we compute the predictions for the considered set of observables $\vec O_{\mathrm{Th}}(\vec p)$. Then, using the experimental information $\vec O_{\mathrm{Exp}}$ available for those observables, we build a likelihood function $\mathcal L(\vec O_{\mathrm{Exp}}|\vec O_{\mathrm{Th}}(\vec p))$ which gives the probability of obtaining the experimental results $\vec O_{\mathrm{\mathrm{Exp}}}$ assuming that the model is correct. The likelihood function $\mathcal L(\vec O_{\mathrm{Exp}}|\vec O_{\mathrm{Th}}(\vec p))$ encodes all the information on how the model is able to reproduce the observed data all over parameter space. Nevertheless, the knowledge of $\mathcal L(\vec O_{\mathrm{Exp}}|\vec O_{\mathrm{Th}}(\vec p))$ in a multidimensional parameter space can be hardly represented and one is led to the problem of reducing that information to one or two-dimensional subspaces. In the profile likelihood method, for each point in the chosen subspace, the highest likelihood over the complementary, marginalized space, is retained. Let us clarify that likelihood -- or chi-squared $\chi^2\equiv -2\log \mathcal L$ -- profiles and derived regions such as the ones we represent, are thus insensitive to the size of the space over which one marginalizes; this would not be the case in a Bayesian analysis, where an integration over the marginalized space is involved. The profile likelihood method seems adequate to our purpose, which is none other than exploring where in parameter space are the different BGL models able to satisfy experimental constraints, without weighting in eventual fine tunings of the models or parameter space volumes. For the numerical computations the libraries GiNaC \cite{Bauer20021} and ROOT \cite{Brun:1997pa}.

\section{Input \label{AP:Input}}

In tables \ref{TAB:AP:MixingMatrices}, \ref{TAB:AP:TreeCharged}, \ref{TAB:AP:TreeNeutral}, \ref{TAB:AP:Loop} and \ref{TAB:AP:Misc} we collect relevant input used in the analysis.

\begin{table}
\begin{center}
\begin{tabular}{|c|c||c|c|}
\hline    $\lambda$ & $0.22535(65)$ & $A$ & $0.811(22)$\\ 
\hline    $\bar{\rho}$ & $0.131(26)$ & $\bar{\eta}$ & $0.345(14)$\\ 
\hline\hline    $\sin^2 \theta_{12}$& $0.320(16)$ & $\sin^2 \theta_{23}$& $0.613(22)$\\ 
\hline    $\sin^2 \theta_{13}$& $0.0246(29)$\\ 
\cline{1-2}
\end{tabular}
\caption{Input for the CKM and PMNS mixing matrices \cite{pdg}.\label{TAB:AP:MixingMatrices}}
\end{center}
\end{table}

\begin{table}
\begin{center}
\begin{tabular}{|c|c||c|c|}
\hline
    $\abs{g_\mu/g_e}^2$ & $1.0018(14)$ & $|g_{RR,\tau \mu}^{S}|$ & $<0.72$\\ 
    $|g_{RR,\tau e}^{S}|$ & $<0.70$ & $|g_{RR,\mu e}^{S}|$ & $<0.035$\\
\hline   
    $\Br(B^+ \to e^+ \nu)$ & $<9.8\cdot 10^{-7}$ & $\Br(D^+_s \to e^+ \nu)$ & $<1.2\cdot 10^{-4}$ \\
    $\Br(B^+ \to \mu^+ \nu)$ & $< 1.0\cdot 10^{-6}$ &  $\Br(D^+_s \to \mu^+ \nu)$ & $5.90(33)\cdot 10^{-3}$\\
    $\Br(B^+ \to \tau^+ \nu)$ & $1.15(23)\cdot 10^{-4}$ &  $\Br(D^+_s \to \tau^+ \nu)$ & $5.43 (31)\cdot 10^{-2}$\\
\hline
    $\Br(D^+ \to e^+ \nu)$ & $<8.8\cdot 10^{-6}$ \\
    $\Br(D^+ \to \mu^+ \nu)$ &  $3.82(33)\cdot 10^{-4}$ \\
    $\Br(D^+ \to \tau^+ \nu)$ & $<1.2\cdot 10^{-3}$\\ 
\hline
    $\frac{\Gamma(\pi^+\to e^+\nu)}{\Gamma(\pi^+\to \mu^+\nu)}$ & $1.230(4)\cdot 10^{-4}$ & $\frac{\Gamma(\tau^-\to \pi^-\nu)}{\Gamma(\pi^+\to \mu^+\nu)}$ & $9703(54)$\\
    $\frac{\Gamma(K^+\to e^+\nu)}{\Gamma(K^+\to \mu^+\nu)}$ & $2.488(12)\cdot 10^{-5}$ & $\frac{\Gamma(\tau^-\to K^-\nu)}{\Gamma(K^+\to \mu^+\nu)}$ &
    $469(7)$\\
\hline
$\frac{\Gamma(B\to D\tau\nu)_{\mathrm{NP}}}{\Gamma(B\to D\tau\nu)_{\mathrm{SM}}}$ & & $\log C$ ($K\to\pi\ell\nu$) & $0.194(11)$\\
\cline{3-4}
$\frac{\Gamma(B\to D^\ast\tau\nu)_{\mathrm{NP}}}{\Gamma(B\to D^\ast\tau\nu)_{\mathrm{SM}}}$ & \\
\cline{1-2}
\end{tabular}
\caption{Constraints on processes mediated at tree level by $H^\pm$ -- section \ref{SEC:ExpConst-sSEC:TreeCharged} --, bounds are given at 95\% CL.\label{TAB:AP:TreeCharged}}
\end{center}
\end{table}

\begin{table}
\begin{center}
\begin{tabular}{|c|c||c|c|}
\hline
    $\Br(\tau^- \to e^-e^-e^+)$ & $<2.7\cdot 10^{-8}$ & $\Br(\tau^- \to \mu^-\mu^-\mu^+)$ & $<2.1\cdot 10^{-8}$\\ 
    $\Br(\tau^- \to e^-e^-\mu^+)$ & $<1.5\cdot 10^{-8}$ & $\Br(\tau^- \to e^-\mu^-e^+)$ & $<1.8\cdot 10^{-8}$\\ 
    $\Br(\tau^- \to \mu^-\mu^-e^+)$ & $<1.7\cdot 10^{-8}$ & $\Br(\tau^- \to \mu^-e^-\mu^+)$ & $<2.7\cdot 10^{-8}$\\ 
\cline{3-4}
    $\Br(\mu^- \to e^-e^-e^+)$ & $<1\cdot 10^{-12}$\\ 
\hline
    $2|M_{12}^K|$ & $<3.5 \cdot 10^{-15}$ \GeV & $2|M_{12}^D|$ & $<9.47\cdot 10^{-15}$ \GeV \\
\cline{3-4}
    $|\epsilon_K|_{NP}\Delta m_K$ & $<7.8 \cdot 10^{-19}$ \GeV \\
\hline
    $\re(\Delta_d)$ & $0.823(143)$ & $\re(\Delta_s)$ & $0.965(133)$ \\
    $\im(\Delta_d)$ & $-0.199(62)$ & $\im(\Delta_s)$ & $0.00(10)$\\
\hline
    $\Br(K_L \to \mu^\pm e^\mp)$ & $<4.7\cdot 10^{-12}$ & $\Br(\pi^0 \to \mu^\pm e^\mp)$ & $<3.6\cdot 10^{-10}$\\
\cline{3-4} 
    $\Br(K_L \to e^-e^+)$ & $<9\cdot 10^{-12}$ \\ 
    $\Br(K_L \to \mu^- \mu^+)$ & $<6.84\cdot 10^{-9}$\\ 
\hline
    $\Br(D^0 \to e^- e^+)$ & $< 7.9\cdot 10^{-8}$ & $\Br(B^0 \to e^+ e^-)$ & $<8.3\cdot 10^{-8}$ \\
    $\Br(D^0 \to \mu^\pm e^\mp)$ & $<2.6\cdot 10^{-7}$ & $\Br(B^0 \to \tau^\pm e^\mp)$ & $<2.8\cdot 10^{-5}$ \\
    $\Br(D^0 \to \mu^- \mu^+)$ & $<1.4\cdot 10^{-7}$ & $\Br(B^0 \to \mu^- \mu^+)$ & $3.6(1.6)\cdot 10^{-10}$ \\
\cline{1-2}
    $\Br(B_s^0 \to e^+ e^-)$ & $<2.8\cdot 10^{-7}$ & $\Br(B^0 \to \tau^\pm \mu^\mp)$ & $<2.2\cdot 10^{-5}$ \\
    $\Br(B_s^0 \to \mu^\pm e^\mp)$ & $<2\cdot 10^{-7}$ & $\Br(B^0 \to \tau^+ \tau^-)$ & $<4.1\cdot 10^{-3}$ \\
\cline{3-4}
    $\Br(B_s^0 \to \mu^- \mu^+)$ & $2.9(0.7)\cdot 10^{-9}$\\
\cline{1-2}
\end{tabular}
\caption{Constraints on processes mediated at tree level by $R$, $I$ -- section \ref{SEC:ExpConst-sSEC:TreeNeutral} --, bounds are given at 95\% CL.\label{TAB:AP:TreeNeutral}}
\end{center}
\end{table}

\begin{table}
\begin{center}
\begin{tabular}{|c|c||c|c|}
\hline    
$\Br(\mu \to e\gamma)$ & $<2.4\cdot 10^{-12}$ & $\Br(B \to X_s \gamma)_{\mathrm{SM}}^{\mathrm{NNLO}}$ & $3.15(23)\cdot 10^{-4}$\\ 
    $\Br(\tau \to e\gamma)$& $<3.3\cdot 10^{-8}$ & $\Br(B \to X_s \gamma)$ & $3.55(35)\cdot 10^{-4}$\\
\cline{3-4}
    $\Br(\tau \to \mu\gamma)$ & $<4.4\cdot 10^{-8}$ \\ 
\hline
    $\Delta T$ & $0.02(11)$ & $F_{Zb\bar b}$ & $<0.0024$ \GeV$^{-1}$ \\
\cline{3-4} 
    $\Delta S$ & $0.00(12)$ \\
\cline{1-2}
\end{tabular}
\caption{Constraints on processes mediated by $R$, $I$, $H^\pm$ at loop level -- section \ref{SEC:ExpConst-sSEC:Loop} --, bounds are given at 95\% CL.\label{TAB:AP:Loop}}
\end{center}
\end{table}

\begin{table}
\begin{center}
\begin{tabular}{|c|c||c|c||c|c|}
\hline $f_{\pi}$ & $0.132(2)$ \GeV & $f_{K}$ & $0.159(2)$ GeV & $f_{D}$ & $0.208(3)$ \GeV\\ 
\hline $f_{D_s}$ & $0.248(3)$ \GeV & $f_{B}$ & $0.189(4)$ GeV & $f_{B_s}$ & $0.225(4)$ \GeV\\ 
\hline $\delta_{\pi^+}$ & $-0.036419(78)$ & $\delta_{K^+}$ & $-0.03580(39)$ & $\delta_{\tau\pi}$ & $0.0016(14)$\\
\hline $\delta_{\tau K}$ & $0.0090(22)$ & $\Delta_{\chi PT}$ & $-3.5(8)\cdot 10^{-3}$ & $f_+^{K\pi}$ & $0.965(10)$\\
\hline
\end{tabular}
\caption{Additional theoretical input -- lattice, radiative corrections -- \cite{lattice,kaon,taucorrections,Pich:2010,Massesrunning}.\label{TAB:AP:Misc}}
\end{center}
\end{table}

\clearpage


\providecommand{\href}[2]{#2}\begingroup\raggedright\endgroup

\end{document}